\documentclass[a4paper,11pt]{article}
\pdfoutput=1

\usepackage{jcappub}

\usepackage[T1]{fontenc}
\usepackage[utf8]{inputenc}

\usepackage{subcaption}
\usepackage{csquotes}
\usepackage{xspace}
\usepackage{float}

\usepackage{multirow}
\usepackage[dvipsnames]{xcolor}
\usepackage{array}
\usepackage{makecell}
\usepackage{hyperref}
\usepackage{amsmath}
\usepackage{mathrsfs}
\usepackage{amssymb}

\newcommand{\lcdm}{$\Lambda$CDM\xspace}
\newcommand{\g}{$g_{\mu\nu}$\xspace}
\newcommand{\f}{$f_{\mu\nu}$\xspace}
\newcommand{\kg}{\kappa_g}
\newcommand{\kf}{\kappa_f}
\newcommand{\mfp}{m_\mathrm{FP}}
\newcommand{\omegal}{\Omega_\Lambda}
\newcommand{\omegade}{\Omega_\mathrm{DE}}

\newcommand{\bic}{$\Delta$BIC\xspace}
\newcommand{\aic}{$\Delta$AIC\xspace}

\title{\boldmath
2D BAO vs 3D BAO: solving the Hubble tension with an alternative cosmological model
}

\author[1,a]{Sowmaydeep Dwivedi,\note{Corresponding author.}}
\author[b]{Marcus Högås}

\affiliation[a]{Department of Physics, Indiana University Bloomington, IN, 47405, USA}
\affiliation[b]{Oskar Klein Centre, Department of Physics, Stockholm University,\\SE-106 91 Stockholm, Sweden}

\emailAdd{sodwive@iu.edu}
\emailAdd{marcus.hogas@fysik.su.se}

\abstract{Ordinary 3D Baryon Acoustic Oscillations (BAO) data are model-dependent, requiring the assumption of a cosmological model to calculate comoving distances during data reduction. Throughout the present-day literature, the assumed model is $\Lambda$CDM. However, it has been pointed out in several recent works that this assumption can be inadequate when analyzing alternative cosmologies, potentially biasing the Hubble constant ($H_0$) low, thus contributing to the Hubble tension. To address this issue, 3D BAO data can be replaced with 2D BAO data, which is only weakly model-dependent. The impact of using 2D BAO data, in combination with alternative cosmological models beyond \lcdm, has been explored for several phenomenological models, showing a promising reduction in the Hubble tension. In this work, we accommodate these models in the theoretically robust framework of bimetric gravity.
This is a modified theory of gravity that exhibits a transition from a (possibly) negative cosmological constant in the early universe to a positive one in the late universe. By combining 2D BAO data with cosmic microwave background and type Ia supernovae data, we find that the inverse distance ladder in this theory yields a Hubble constant of $H_0 = (71.0 \pm 0.9) \, \mathrm{km/s/Mpc}$, consistent with the SH0ES local distance ladder measurement of $H_0 = (73.0 \pm 1.0) \, \mathrm{km/s/Mpc}$.
Replacing 2D BAO with 3D BAO results in $H_0 = (68.6 \pm 0.5) \, \mathrm{km/s/Mpc}$ from the inverse distance ladder. Thus, the choice of BAO data significantly impacts the Hubble tension, with ordinary 3D BAO data exacerbating the tension, while 2D BAO data provides results consistent with the local distance ladder.
}

\begin{document}
\maketitle
\flushbottom

\section{Introduction}
\label{sec:intro}
In the early 20th century, the advent of Einstein's general theory of relativity (GR) revolutionized our understanding of gravity, providing a robust framework that successfully describes a myriad of cosmic phenomena---ranging from the perihelion precession of Mercury to gravitational lensing and the existence of black holes \cite{Will:2014kxa}. This, together with the Standard Model of particle physics, has paved the path for the {$\Lambda$} Cold Dark Matter (\lcdm) model, providing a cosmological framework that accounts for the vast majority of current observations with astonishing precision.

However, as we have entered the era of high-precision cosmology, a handful of tensions has emerged. The most prominent is the Hubble tension which refers to the fact that the local measurement of the Hubble constant significantly exceeds the inferred value from the inverse distance ladder.
The most discrepant estimates is between the local $H_0$ measurement from the SH0ES team, $H_0 = (73.0 \pm 1.0) \, \mathrm{km/s/Mpc}$ \cite{Riess:2021jrx}, and the inferred value from the inverse distance ladder using \emph{Planck} satellite data, $H_0 = (67.8 \pm 0.5) \, \mathrm{km/s/Mpc}$ \cite{Planck:2018vyg}. 
This amounts to a $5 \, \sigma$ discrepancy, making it difficult to explain as a mere statistical fluke. Despite diligent searches for possible systematic errors to explain the tension, it has just been increasing during the last decade.

Accordingly, there is an intense discussion in the contemporary literature whether new physics can alleviate the tension. That is, lowering the local distance ladder (SH0ES) value or increasing the inverse distance ladder value.\footnote{A combination of the two is of course also an option. See e.g. \cite{Vagnozzi:2023nrq}.}
The former can be achieved by postulating a new gravitational degree of freedom that modifies gravity on galactic and astronomical scales---a fifth force. This results in a recalibration of the cosmic distance ladder, changing the local distance ladder value of $H_0$   \cite{Desmond:2019ygn,Hogas:2023pjz,Hogas:2023vim}.
On the other hand, the inferred value of $H_0$ from the inverse distance ladder assumes a \lcdm cosmological model.
That is, alternative expansion histories can change the inferred value of $H_0$ which may result in increased consistency with the local distance ladder. This can be achieved by introducing new particles, changing the properties of dark matter or dark energy, or by modifying the theory of gravity itself, see Refs.~\cite{Knox:2019rjx,Vagnozzi:2019ezj,DiValentino:2021izs,Schoneberg:2021qvd,Abdalla:2022yfr} for some examples.
The focus of this paper is on the latter option, and more precisely on the bimetric theory of gravity and its effect on the Hubble tension.

However, the combination of standard 3D BAO, cosmic microwave background (CMB), and type Ia supernovae (SNIa) data, tightly restricts any deviation from a \lcdm expansion history at late times (redshift $z \lesssim 2$), leaving little room for a late-time alternative expansion history to resolve the Hubble tension.
On the other hand, several recent studies emphasize that the 3D BAO data reduction inherently assumes a cosmological model, universally assumed to be \lcdm or some slight variation thereof \cite{Carter:2019ulk,Sanz-Wuhl:2024uvi,DESI:2024ude}. This introduces a model dependence that is incompatible with alternative expansion histories, raising concerns about the appropriateness of using 3D BAO data to restrict non-standard cosmological models, see for example Refs.~\cite{Carvalho:2015ica,Anselmi:2018vjz,ODwyer:2019rvi,DiValentino:2019ffd,Camarena:2019rmj,Nunes:2020hzy,Nunes:2020uex,Anselmi:2022exn,Bernui:2023byc,Gomez-Valent:2023uof,Favale:2024sdq,Ruchika:2024lgi}. In particular, the \lcdm assumption may bias the Hubble constant low, thereby giving rise to the Hubble tension.

To address this issue, the 2D transverse BAO scale (2D BAO) can be used, as it is only weakly model-dependent  \cite{Sanchez:2010zg,Carvalho:2015ica,Alcaniz:2016ryy,deCarvalho:2017xye,Carvalho:2017tuu,Nunes:2020hzy,deCarvalho:2021azj}. As opposed to 3D BAO, 2D BAO allows for significant deviations from \lcdm at late times. Recent studies suggest that 2D BAO, combined with a modified late-time expansion history, favours a Hubble constant consistent with the SH0ES measurement, thus alleviating the tension \cite{Li:2020ybr,Yang:2021eud,Akarsu:2021fol,Akarsu:2022typ,Akarsu:2023mfb,Gomez-Valent:2024tdb,Hernandez-Almada:2024ost}. These results are based on phenomenological models such as the $\Lambda_s$CDM model 
\cite{Akarsu:2021fol,Akarsu:2022typ,Akarsu:2023mfb} and phenomenological emergent dark energy \cite{Li:2019yem}. In the current paper, we analyze the cosmology of bimetric gravity which is a theoretically robust framework that accommodates a range of these phenomenological models. Thus, we are not surprised to find that this theory has beneficial effects on the Hubble tension when 2D BAO data is used in combination with CMB and SNIa.

Bimetric gravity is a natural extension of GR, exhibiting a massive spin-2 field in addition to the massless spin-2 field \cite{S_F_Hassan_2012,Hassan:2012wr}. The theory is observationally viable, as demonstrated in a number of papers \cite{vonStrauss:2011mq,Sjors:2011iv,Akrami:2012vf,Enander:2013kza,Babichev:2013pfa,Koennig:2013fdo,Enander:2015kda,Max:2017flc,Dhawan:2017leu,Platscher:2018voh,Luben:2018ekw,Hogas:2019ywm,Luben:2020xll,Lindner:2020eez,Caravano:2021aum,Hogas:2021fmr,Hogas:2021lns,Hogas:2021saw,Hogas:2022owf}. 
Among its virtues is the existence of self-accelerating cosmological solutions where the accelerated expansion of the Universe is a result of the interaction between the two spin-2 fields---no cosmological constant is needed \cite{Volkov:2011an,vonStrauss:2011mq,Comelli:2011zm,Volkov:2012wp,Volkov:2012zb,Akrami:2012vf,Volkov:2013roa,Koennig:2013fdo,Hogas:2021fmr}. Another interesting feature is that, under certain conditions, the massive spin-2 field provides a dark matter particle \cite{Aoki:2016zgp,Babichev:2016bxi,Babichev:2016hir}. 
The theory is also scientifically tractable in the sense that even the most general version of the theory only introduces four additional theory parameters, which can be constrained observationally.

Nevertheless, bimetric gravity provides a rich spectrum of cosmological expansion histories that can modify the expansion rate both pre- and post-recombination, thus changing the $H_0$ value inferred from the inverse distance ladder. The effect of bimetric gravity on the Hubble tension was investigated in Ref.~\cite{Mortsell:2018mfj} for a restricted subclass of models---more precisely, for some special two-parameter models. It was shown that, for this subclass of models, the tension is eased only very slightly.

In the present work however, we do not restrict ourselves to a limited type of submodel but analyze the most general bimetric model. We investigate the effects of combining different data sets and their impact on the Hubble tension. Specifically, we compare the results using ordinary 3D BAO data with the results using transverse 2D BAO data (BAOtr). 
We infer $H_0 = (71.0 \pm 0.9) \, \mathrm{km/s/Mpc}$ from the inverse distance ladder when the general bimetric model is fitted with data from the cosmic microwave background, type Ia supernovae, and 2D BAO. This is on the $2\, \sigma$ border of the SH0ES value for $H_0$.
Using 3D BAO instead of 2D BAO, the inferred value is $H_0 = (68.6 \pm 0.5) \, \mathrm{km/s/Mpc}$, representing only a slight ease of the tension, being in the $4.4 \, \sigma$ tail of the SH0ES value. The disparate result of 2D BAO compared with 3D BAO suggests that the cosmological model dependence in ordinary 3D BAO data may bias $H_0$ to a low value.

\paragraph{Notation.} Unless stated otherwise, we use geometrized units where the speed of light and Newton's gravitational constant are unity, $c = G = 1$. In this case, length, time, and mass all have the same units $\mathsf{L} = \mathsf{T} = \mathsf{M}$. Geometrical quantities pertaining to the second metric \f are denoted by tildes---if not, then pertaining to the physical metric \g. Derivatives with respect to time are denoted by an overdot, so for example $\dot{a} = \frac{da}{dt}$.
We let $\omega_x = \Omega_{x,0} h^2$ where $h = H_0 / (100 \, \mathrm{km/s/Mpc})$ is the normalized Hubble constant and $\Omega_{x,0}$ denotes the present-day density of the species $x$. The Hubble constant $H_0$ is given in units of $\mathrm{km/s/Mpc}$.

\section{Bimetric gravity}
\subsection{Historical background}
Bimetric gravity posits the existence of two dynamical spin-2 fields, or metrics, governing gravitational interactions. The first steps towards the present-day, ghost-free, formulation of this theory was taken by Fierz and Pauli in 1939 \cite{fierz1939relativistic}. They formulated a consistent linearized theory for a freely propagating massive spin-2 field in Minkowski space-time. However, in 1972 Boulware and Deser examined the coherence of a broad range of nonlinear extensions of this theory. Their conclusion was that the introduction of an additional propagating ghost-like scalar mode is unavoidable in any nonlinear extension of the theory \cite{boulware1972can}.
Nevertheless, building upon works by de Rham, Gabadadze, and Tolley \cite{Gabadadze_2009,de_Rham_2010,DERHAM2010334,PhysRevD.82.044020,de_Rham_2011,Hassan_2011}, in 2011 Hassan and Rosen formulated the ghost-free version of bimetric gravity with two dynamical metrics (spin-2 fields) \cite{S_F_Hassan_2012,Hassan:2011ea,Hassan:2018mbl}. This marked the rebirth of massive gravity which has been intensely studied since.

\subsection{Theory}
The ghost-free action for bimetric gravity reads
\begin{equation}\label{HR_action}
\begin{split}
\mathcal{S
}_\mathrm{HR}  = \int d^4x & \left[ \frac{1}{2\kg} \sqrt{-\det g} \, R + \frac{1}{2\kf}\sqrt{-\det f} \, \widetilde{R} - \sqrt{-\det g} \sum_{n=0}^4\beta_ne_n(S) + \right. \\ & \left. + \sqrt{-\det g} \, \mathcal{L}_m + \sqrt{-\det f} \, \widetilde{\mathcal{L}}_m \right] .
\end{split}
\end{equation}
In the action, \g and \f are the two metrics (spin-2 fields) and $R$ and $\widetilde{R}$ are the corresponding Ricci scalars. The two metrics are dynamical with each metric exhibiting an Einstein--Hilbert term, $\kg = 8 \pi G/c^4$ represents the gravitational constant for \g, while $\kf$ denotes the gravitational constant for \f. The Lagrangians $\mathcal{L}_m$ and $\widetilde{\mathcal{L}}_m$ characterize two independent matter sectors coupled to $g_{\mu\nu}$ and $f_{\mu\nu}$, respectively \cite{Rham_2015,de_Rham_2014}. For simplicity, we only consider matter fields coupled to \g (i.e., assuming $\widetilde{\mathcal{L
}}_m =0$), which we accordingly identify as the physical metric, determining the geodesics of freely falling observers.
Further, the elementary symmetric polynomials of the square root ($S$) of the two metrics, denoted by $e_n(S)$ in \eqref{HR_action}, are, 
\begin{equation}\label{ESP}
\begin{split}
e_0(S) &= 1 \,,
\qquad
e_1(S) = [S]\,,
\qquad
e_2(S) = \frac{1}{2}([S]^2-[S^2])\,,
\\
e_3(S) &= \frac{1}{6}([S]^3-3[S^2][S]+2[S^3]) \,,
\qquad
e_4(S) = \det(S).
\end{split}
\end{equation}
In \eqref{ESP}, $[S] =$ Tr $S$. The square root of the two metrics, $S$, is defined by the equation ${S^\mu}_\rho {S^\rho}_\nu = g^{\mu\rho}f_{\rho\nu}$ \cite{Hassan_2018,higham2008functions}.\footnote{Note that this equation has no unique solution generically. However, in Ref.~\cite{Hassan_2018} it was argued that the principal square root is the appropriate solution as it guarantees a sensible space-time interpretation of the theory.}
The five coefficients $\beta_n$ of the polynomials $e_n(S)$ are constants with the dimension of curvature $1 / \mathsf{L}^2$ with $\beta_1,\beta_2$, and $\beta_3$ determining the interaction between the two metrics and $\beta_0$ and $\beta_4$ contributing with a cosmological constant to the $g$-metric and $f$-metric, respectively.

As opposed to some other modified gravity theories such as Horndeski theory \cite{Horndeski:1974wa}, bimetric gravity features a finite number of theory parameters which makes the theory scientifically tractable, with the possibility to falsify the theory and constrain the theory parameters. 
As they stand the $\beta$-parameters cannot be constrained by observations due to their invariance under the rescaling
\begin{equation}
\label{eq:rescaling}
    (f_{\mu\nu} , \kf , \beta_n) \to (\omega f_{\mu\nu} , \omega \kf , \omega^{-n/2} \beta_n),
\end{equation}
with $\omega$ being an arbitrary constant. Therefore, rescaling-invariant parameters are introduced \cite{Luben:2020xll,Hogas:2021fmr} as 
\begin{equation}\label{B_param}
    B_n \equiv \frac{\kappa_g\beta_nc^n}{H^2_0} .
\end{equation}
Here, $c$ is the proportionality constant between the metrics in the final de Sitter phase in the cosmological infinite future where $f_{\mu \nu} = c^2 g_{\mu \nu}$. For more details, see Refs.~\cite{Hogas:2021fmr,Hogas:2022owf}. The Hubble constant $H_0$ is included in the definition of $B_n$ to render the $B_n$-parameters dimensionless. Due to the $B_n$-parameters being invariant under the rescaling, they can be observationally constrained.

To enable an intuitive interpretation, we reparameterize from $(B_0 , B_1 , B_2, B_3, B_4)$ to what we refer to as the physical parameters $(\theta,\mfp,\omegal,\alpha,\beta)$, defined by,
\begin{subequations}\label{eq:y}
\begin{align}
\label{eq:y:1}
\tan^2\theta &= \frac{B_1 + 3B_2 + 3B_3 + B_4}{B_0 + 3B_1 + 3B_2 + B_3},
\\
\label{eq:y:2}
\mfp^2 &= \frac{B_1 + 2B_2 + B_3}{\sin^2 \theta},
\\
\label{eq:y:3}
\omegal &= \frac{B_0}{3}+ B_1 + B_2 +\frac{B_3}{3},\\
\label{eq:y:4}
\alpha &= -\frac{B_2+B_3}{B_1+2B_2+B_3},\\
\label{eq:y:5}
\beta &= \frac{B_3}{B_1+2B_2+B_3}.
\end{align}    
\end{subequations}
As implied by their name, the physical parameters carry specific interpretations. Firstly, the mixing angle $\theta\in [0,\pi/2]$ governs the mixing of the massive and massless spin-2 fields, also known as mass eigenstates. In the limit $\theta \to 0$, GR is retained whereas in the limit $\theta \to \pi /2$, dRGT massive gravity is retained. The parameter $\mfp$ represents the mass of the massive spin-2 field. The effective cosmological constant in the final de Sitter phase in the Universe's expansion history is denoted by $\omegal$. Lastly, the parameters $\alpha$ and $\beta$ play a crucial role in determining the Vainshtein screening mechanism that is responsible for recovering GR results on solar-system scales.

\subsection{Cosmology}
From the Hassan--Rosen action \eqref{HR_action}, one can derive the Friedmann equation governing the evolution of a homogeneous and isotropic universe, as a function of the redshift $z$,
\begin{equation}
\label{eq:Friedmann}
    \left( \frac{H(z)}{H_0} \right)^2 = \Omega_m(z) + \Omega_r(z) + \omegade(z).
\end{equation}
Thus, $\Omega_i = \rho_i / \rho_c$ denotes the dimensionless energy density of species $i$, that is, the physical energy density $\rho_i$ measured in units if the present-day critical density $\rho_c = 3 H_0^2 / \kg$.
Here, we have set the spatial curvature to zero, $\Omega_k = 0$. $H = \dot{a}/a$ is the Hubble parameter with $a$ being the scale factor of the physical metric and $H_0$ is the present-day value of $H$, that is, the Hubble constant. Further, $\Omega_m$ and $\Omega_r$ denote the energy density of matter and radiation, respectively, both measured in units of the present-day critical energy density. What distinguishes eq.~\eqref{eq:Friedmann} from the ordinary Friedmann equation of the \lcdm model is the last term, $\omegade$, which results from the interaction between the two metrics. It reads,
\begin{equation}\label{DE}
    \omegade(z) = \omegal - \sin^2 \theta \, \mfp^2 \left[ 1-y(z)\right] \left[ 1 + \alpha(1-y(z)) + \frac{\beta}{3} (1-y(z))^2\right],
\end{equation}
where $y \equiv \widetilde{a}/a$ is the ratio of the scale factors of the two metrics. The first term, $\omegal$, gives a cosmological constant contribution whereas the remaining terms give a dynamical contribution that evolves with the redshift $z$.  

From the conservation equations of matter and radiation it follows that,
\begin{equation}\label{Omega}
    \Omega_m(z) = \Omega_{m,0} (1+z)^3, \quad \Omega_r(z) = \Omega_{r,0} (1+z)^4,
\end{equation}
with $\Omega_{m,0}$ and $\Omega_{r,0}$ being the present-day matter density and radiation density, respectively. Finally, an equation can be derived for $y$ in form of a quartic polynomial in $y$. The full expression is shown in eq.~\eqref{eq:yPoly}.

With $y$ being a solution to a quartic polynomial, there are up to four real solutions. Here, we choose the solution which is commonly referred to as the ``finite branch''. This branch guarantees the absence of the Higuchi ghost and allows for a screening mechanism that restores GR on solar-system scales \cite{Higuchi:1986py,Fasiello:2013woa,DeFelice:2014nja,Konnig:2015lfa,Hogas:2021fmr}. This branch has a finite range in $y$, starting with $y=0$ at the Big Bang, then monotonically increasing until $y=1$ is reached in the infinite future, that is, in the final de Sitter phase.

In the early-time limit $z \to \infty$, one can show that the bimetric contribution to the Friedmann equation, $\omegade$, generically assumes the form of a cosmological constant with magnitude
\begin{equation}
\label{eq:OmegaDEearly}
    \omegade |_{z \to \infty} = \omegal - \sin^2 \theta \, \mfp^2 \left(  1 + \alpha + \frac{\beta}{3} \right)  .
\end{equation}
That is, at early times the background evolution of the Universe is according to a \lcdm model with a value of the, possibly negative, cosmological constant that is given by eq.~\eqref{eq:OmegaDEearly} \cite{Hogas:2021fmr}. In the infinite future, the Universe approaches a de Sitter phase where the cosmological constant is set by $\omegal$ \cite{Hogas:2021fmr}. So, bimetric cosmology exhibits two \lcdm phases with generically different cosmological constants, one in the early universe and one in the the late universe.
The redshift ranges of these phases are primarily set by the mass of the massive spin-2 field, $\mfp$. In the transition between the two \lcdm phases, there is a rich spectrum of expansion histories depending on the physical parameters \eqref{eq:y}. One can show that if $\omegade$ is negative, the equation of state $w_\mathrm{DE} \geq -1$ whereas if $\omegade$ is positive the equation of state is $w_\mathrm{DE} \leq -1$.\footnote{For a more general discussion of a dark energy fluid with such properties, see Ref.~\cite{Ozulker:2022slu}.} In other words, $\omegade$ grows with time, thus presenting a phantom dark energy contribution to the Friedmann equation \eqref{eq:Friedmann}. However, $w_\mathrm{DE
} \to -1$ fast enough in the late universe, thus avoiding a Big Rip \cite{Mortsell:2017fog,Hogas:2021fmr}. 

To enable a statistical data analysis, we must solve the Hubble parameter as a function of redshift $H(z)$. This is done by solving the cosmological equations of motion numerically, that is, eqs.~\eqref{eq:Friedmann}, \eqref{DE}, \eqref{Omega}, \eqref{eq:yPoly}.
The value of $y$ at $z=0$ is determined by solving eq.~\eqref{eq:ycub} numerically, choosing the finite branch solution with $y_0$ in the range $0 \leq y_0 \leq 1$.
Subsequently, this value of $y_0$ allows the calculation of $\Omega_\mathrm{DE,0}$ using eq.~\eqref{DE}. Further, the application of Friedman's equation \eqref{eq:Friedmann} at $z=0$ yields
\begin{equation}\label{OM}
    \Omega_{m,0} = 1 - \Omega_\mathrm{DE,0} - \Omega_{r,0},
\end{equation}
since $\left. \frac{H^2(z)}{H_0^2} \right|_{z=0} = 1$, by definition.
With $\Omega_{m,0}$ and $\Omega_{r,0}$ being determined,\footnote{In Section~\ref{sec:CMB}, we show how $\Omega_{r,0}$ is determined from the CMB temperature.} the remaining step to obtain $H(z)$ is to solve eq.~\eqref{eq:yPoly} for $y$ at each redshift, plugging the solution into eq.~\eqref{DE} to determine $\omegade(z)$ and then finally $H(z)$ via eq.~\eqref{eq:Friedmann}. 

\section{Methodology and data}
\subsection[The SH0ES H0 estimate]{The SH0ES $H_0$ estimate}
The SH0ES team estimate of $H_0$ is based on a three-rung cosmic distance ladder \cite{Riess:2021jrx}. In the first rung, the period-luminosity relation (PLR) for Cepheid variable stars \cite{Leavitt:1912zz} is calibrated in three anchor galaxies---the Milky Way (MW), NGC 4258, and the Large Magellanic Cloud (LMC). To calibrate the PLR, the distance to these Cepheids must be known.
In the MW, the distances can be estimated geometrically from parallax measurements \cite{Riess:2018byc,Riess_2021}.
The distance to NGC 4258 is estimated from water masers close to the centre of this galaxy \cite{Reid:2019tiq}.
The distance to the LMC is estimated from observations of detached eclipsing binaries \cite{Pietrzynski:2019cuz}. 

In the second rung, the absolute peak magnitude of type Ia supernovae is calibrated in galaxies hosting both SNIa and Cepheids.
In the third rung, SNIa are observed in the Hubble flow which yields the magnitude-redshift relation \cite{Scolnic:2021amr}. With the calibrated value for the absolute SNIa peak magnitude, $H_0$ is finally obtained from the intercept of the magnitude-redshift relation. The 2022 baseline result from the SH0ES team, fitting all three rungs simultaneously, is $H_0 = (73.0 \pm 1.0) \, \mathrm{km/s/Mpc}$ \cite{Riess:2021jrx}.
This can be used as a prior on $H_0$ in which case the contribution to the likelihood is given by
\begin{equation}
    -2 \ln \mathcal{L}_\mathrm{SH0ES} = \left(\frac{H_0^\mathrm{model} - 73.0 \, \mathrm{km/s/Mpc}}{1.0 \, \mathrm{km/s/Mpc}}\right)^2.
\end{equation}

\subsection{Cosmic Microwave Background}
\label{sec:CMB}
As the temperature of the Universe had dropped sufficiently, neutral hydrogen was formed and the photons decoupled from the baryon-photon fluid and began to free stream. This is known as recombination, or photon decoupling, and happened at a redshift of $z_* \simeq 1090$. Today, we can observe these photons as the cosmic microwave background radiation, carrying the imprint of the fluctuations in the baryon-photon fluid at the last-scattering surface at redshift $z_*$.

The \emph{Planck} satellite observations of the temperature fluctuations in the CMB can be used to estimate $H_0$ assuming a model, for example \lcdm \cite{Planck:2018vyg}. However, this assumes a \lcdm cosmology. Here, we investigate the inferred value of $H_0$ in the case of bimetric cosmology.
For this purpose, we use the CMB compressed likelihood featuring the three parameters $(\mathcal{R},l_A,\omega_b)$. The ``shift parameter'' $\mathcal{R}$ encodes the angular scale of the Hubble horizon at the photon decoupling epoch and is defined by \cite{Efstathiou:1998xx}
\begin{equation}
    \mathcal{R} = \sqrt{\Omega_{m,0} H_0^2} \, D_A(z_*).
\end{equation}
The comoving angular diameter distance $D_A$ is calculated by
\begin{equation}
    D_A(z) = \int_{0}^{z} \frac{dz}{H(z)},
\end{equation}
assuming a spatially flat cosmology. The redshift $z_*$ of photon decoupling is given by the analytical approximation in Ref.~\cite{Hu:1995en}.

The sound horizon at photon decoupling $r_s(z_*)$ defines a standard ruler which is imprinted in the CMB temperature fluctuations. This manifests itself as a maximal correlation between hotspots in the CMB temperature map occurring at the angular scale $\theta_\mathrm{CMB}$, given by
\begin{equation}
\label{eq:thetaCMB}
    \theta_\mathrm{CMB} = r_s(z_*) / D_A(z_*).
\end{equation}
The parameter $l_A$ is the multipole number corresponding to this angular scale,
\begin{equation}
\label{eq:lA}
    l_A = \pi D_A(z_*) / r_s(z_*),
\end{equation}
where $r_s$ is the comoving sound horizon,
\begin{equation}
    r_s(z) = \int_{z}^{\infty} \frac{c_s(z)}{H(z)} dz,
\end{equation}
and $c_s$ is the sound speed,
\begin{equation}
    c_s(z) = \frac{1}{\sqrt{3 \left( 1 + \frac{3 \omega_b}{4 \omega_\gamma} \frac{1}{1+z}\right) }}.
\end{equation}
The photon energy density $\omega_\gamma$ can be determined from the CMB temperature via the relation \cite{Chen:2018dbv}
\begin{equation}
    \label{eq:omegar}
    \frac{3}{4\omega_\gamma} = 31500 \, \left( \frac{2.7 \, \mathrm{K}}{T_\mathrm{CMB}}\right)^4, \quad T_\mathrm{CMB} = 2.7255 \, \mathrm{K}.
\end{equation}
The present-day radiation density is set by,
\begin{equation}
    \label{eq:Omegar0}
    \Omega_{r,0} = \Omega_{\gamma,0} (1 + 0.2271 N_\mathrm{eff}),
\end{equation}
with $N_\mathrm{eff}$ being the effective number of neutrino species, which we set to $N_\mathrm{eff} = 3.046$ \cite{Planck:2018vyg}.

Finally,
\begin{equation}
    \omega_b = \Omega_{b,0} h^2,
\end{equation}
encodes the present-day baryon density. 

The latest \emph{Planck} values for $(\mathcal{R},l_A,\omega_b)$ are given in Tab.~\ref{tab:PlanckData} and the CMB likelihood is calculated as,
\begin{equation}
    -2 \ln \mathcal{L}_\mathrm{CMB} = \sum_{ij} \Delta_i C^{-1}_{ij} \Delta_j.
\end{equation}
Here, $\Delta$ is the difference between the model prediction and the observed values,
\begin{equation}
    \Delta^T = \left(\mathcal{R}^\mathrm{model}-\mathcal{R}^\mathrm{obs} , l_A^\mathrm{model} - l_A^\mathrm{obs} , \omega_b^\mathrm{model} - \omega_b^\mathrm{obs}\right),
\end{equation}
$C^{-1}$ is the inverse covariance matrix and the covariance matrix $C$ can be read off from the correlation matrix $\rho$ via the relation $C_{ij} = \sigma_i \sigma_j \rho_{ij}$ (no summation implied) with $\sigma_i$ and $\rho_{ij}$ given in Tab.~\ref{tab:PlanckData}.

\begin{table}
  \centering
  \renewcommand*{\arraystretch}{1.4}
  \begin{tabular}{c|c|ccc}
    \hline \hline
     & \emph{Planck} & $\mathcal{R}$ & $l_A$ & $\omega_b$\\

    \hline
    
    $\mathcal{R}$ & $1.7493^{+0.0046}_{-0.0047}$ & $1.0$ & $0.47$ & $-0.66$\\

    $l_A$ & $301.462^{+0.089}_{-0.090}$ & $0.47$ & $1.0$ & $-0.34$\\

    $\omega_b$ & $0.02239 \pm 0.00015$ & $-0.66$ & $-0.34$ & $1.0$\\

\hline \hline
  \end{tabular}
  \caption{The $68 \, \%$ CL limits on the CMB compressed likelihood \cite{Chen:2018dbv}. The last three columns show the correlation matrix $\rho_{ij}$.}
  \label{tab:PlanckData}
\end{table}

\subsection{Type Ia supernovae}
Type Ia supernovae are standardizable candles and can thus be used to probe the expansion history of the Universe. Here, we use the Pantheon+ data sample \cite{Scolnic:2021amr}, probing the peak apparent magnitude $m_B$ of 1701 SNIa light curves in the redshift range $0.001 \leq z \leq 2.26$. To retain the supernovae in the Hubble flow, we use only those with a redshift greater than 0.023.\footnote{The data set is available at \url{https://github.com/PantheonPlusSH0ES/DataRelease/tree/main/Pantheon\%2B_Data}, last checked 2 July 2024.}
The model prediction of the peak $B$-band magnitude is given by,
\begin{equation}
    m_B(z) = \mathcal{M} + 5 \log_{10} \mathcal{D}_L(z),
\end{equation}
where we marginalize over the intercept $\mathcal{M}$ when calculating the likelihood. For details, see for example Refs.~\cite{Goliath:2001af,Hogas:2022owf}.
Further, the dimensionless luminosity distance $\mathcal{D}_L$ is defined as
\begin{equation}
    \mathcal{D}_L(z) = (1+z) \int_0^z \frac{dz}{E(z)}
\end{equation}
and $E(z)$ is the normalized expansion rate, $E(z) = H(z) / H_0$.

\subsection{Baryon Acoustic Oscillations}
The sound horizon of the baryon-photon fluid is not only imprinted in the CMB temperature fluctuations but also in the large-scale structure of matter (baryons). These fluctuations in the matter distribution are known as baryon acoustic oscillations and the angular scale $\theta_\mathrm{BAO}$ corresponding to the typical angular (transverse) separation of galaxies on the sky, at some redshift $z$, is given by,
\begin{equation}
\label{eq:thetaBAO}
    \theta_\mathrm{BAO}(z) = r_s(z_d) / D_A(z).
\end{equation}
Here, $r_s(z_d)$ is the comoving sound horizon at the baryon drag epoch which occurred when the baryons were released from the Compton drag of the photons. Due to the excess of photons over baryons, this occurred at a slightly later point in time than the photon decoupling. The redshift of the drag epoch $z_d$ is calculated using the analytical approximation in Ref.~\cite{Hu:1995en} and is $z_d \simeq 1060$.

In the standard 3D BAO data reduction, one infers for example $D_V(z) / r_s(z_d)$. That is, the ratio of the volume-average distance and the sound horizon at the baryon drag epoch. The volume-average distance is defined as
\begin{equation}
    D_V(z) =  \sqrt[3]{D_A^2(z) \frac{z}{H(z)}}.
\end{equation}
One can identify the two factors of $D_A(z)$ as angular distances and $z / H(z)$ as a radial distance. To infer $D_V(z) / r_s(z_d)$ from BAO data, one must calculate the 3D fiducial comoving distances, including the radial distance \cite{Carvalho:2015ica,Ruchika:2024lgi}. For the latter, a  cosmological model must be assumed, and the universal assumption in the literature is a \lcdm model. Due to the assumption of a \lcdm cosmological model in the BAO data reduction, in recent years it has been questioned whether it is appropriate to use ordinary 3D BAO data when analyzing alternative cosmological models, see for example Refs.~\cite{Anselmi:2018vjz,ODwyer:2019rvi,DiValentino:2019ffd,Camarena:2019rmj,Nunes:2020hzy,Nunes:2020uex,Anselmi:2022exn,Bernui:2023byc,Gomez-Valent:2023uof,Favale:2024sdq}. In particular, the \lcdm assumption may bias the Hubble constant low when considering alternative cosmological models. 

To circumvent this issue, one can use the 2D transversal BAO scale (BAOtr) which can be obtained without assuming any fiducial cosmology. Here instead, in the BAO data reduction one calculates the 2-point angular correlation function in thin, non-overlapping, redshift shells to infer $\theta_\mathrm{BAO}(z)$ at a set of redshifts.
The weak model dependence that remains is due to corrections for projection effects, which is however minimized by choosing thin redshift bins (see for example Ref.~\cite{Sanchez:2010zg}).
The price to pay for minimizing the model-dependence is that the errors grow by roughly one order of magnitude, from $\sim 1 \, \%$ to $\sim 10 \, \%$.

Here, we use 15 measurements of the angular BAO scale $\theta_\mathrm{BAO}$ in the redshift range $0.11 \leq z \leq 2.225$, obtained from the Sloan Digital Sky Survey (SDSS) data releases DR7, DR10, DR11, DR12, and DR12Q and compiled in Tab.~\ref{tab:BAOtr}. The likelihood is calculated as,
\begin{equation}
    -2 \ln \mathcal{L}_\mathrm{BAOtr} = \sum_{i=1}^{15} \left( \frac{\theta_\mathrm{BAO}^\mathrm{model}(z_i) - \theta_\mathrm{BAO}^\mathrm{obs}(z_i)}{\sigma_i}\right)^2.
\end{equation}
When using ordinary 3D BAO data to compare with the BAOtr results, we use the data from the SDSS in combination with the Dark Energy Spectroscopic Instrument (DESI) \cite{DESI:2024mwx}. 

\begin{table}
  \centering
  \begin{tabular}{c|ccccccccccccccc}
    \hline \hline
     $z$ & 0.11 & 0.236 & 0.365 & 0.45 & 0.47 & 0.49 & 0.51 & 0.53\\

     $\theta_\mathrm{BAO}$ (deg) & 19.8 & 9.06 & 6.33 & 4.77 & 5.02 & 4.99 & 4.81 & 4.29\\

     $\sigma_\mathrm{BAO}$ (deg) & 3.26 & 0.23 & 0.22 & 0.17 & 0.25 & 0.21 & 0.17 & 0.30 \\[2mm]
    \hline
     \\[-3mm]
     
     $z$ & 0.55 & 0.57 & 0.59 & 0.61 & 0.63 & 0.65 & 2.225\\

     $\theta_\mathrm{BAO}$ (deg) & 4.25 & 4.59 & 4.39 & 3.85 & 3.90 & 3.55 & 1.77\\

     $\sigma_\mathrm{BAO}$ (deg) & 0.25 & 0.36 & 0.33 & 0.31 & 0.43 & 0.16 & 0.31\\

\hline \hline
  \end{tabular}
  \caption{Transversal BAO data (BAOtr). Adopted from Ref.~\cite{Nunes:2020hzy}, compiling data points from Refs.~\cite{Carvalho:2015ica,Alcaniz:2016ryy,deCarvalho:2017xye,Carvalho:2017tuu,deCarvalho:2021azj}.}
  \label{tab:BAOtr}
\end{table}

\subsection{Consistency constraints}
\label{sec:ConsConstr}
There are some regions in the bimetric parameter space $(\theta,\mfp,\omegal,\alpha,\beta)$ that must be avoided \cite{Hogas:2021fmr}. First, to have a continuous, real-valued cosmology devoid of the Higuchi ghost, we must restrict the range of possible values of the physical parameters. Second, observations require GR results to be restored on solar-system scales. Thus, there must exist a screening mechanism hiding the extra degrees of freedom on these scales. To ensure the existence of such a mechanism, we must impose additional constraints on the parameter space. Those are presented in Ref.~\cite{Hogas:2021fmr} and we refer to these, together with the restrictions imposed by the Higuchi bound, as consistency constraints.

Bimetric submodels are usually defined by setting one or several of the $B$-parameters (or, equivalently, $\beta$-parameters) to zero. We note that the consistency constraints presented here, together, require that $B_1$, $B_2$, and $B_3$ are all non-zero \cite{Hogas:2021fmr}.\footnote{The $B$-parameters can be expressed in terms of the physical parameters by inverting eq.~\eqref{eq:y}.} Thus, the minimal submodel consistent with these constraints is the $B_1 B_2 B_3$-model which is the reason why we do not consider submodels with fewer parameters.
In the MCMC sampling, we set the likelihood to zero at the points where the consistency constraints are violated.

\subsection{MCMC sampling}
Markov Chain Monte Carlo (MCMC) methods can be employed in sampling complex probability distributions. Its application is particularly prevalent in Bayesian statistics and computational physics. 
MCMC operates on the principle of constructing a Markov Chain, where each state in the chain represents a possible configuration of model parameters. The transition from one state to the next is governed by a Markov process, ensuring that the next state depends only on the current state. Over time, the chain converges to a stationary distribution, and samples drawn from this distribution provide an approximation of the posterior distribution.
We have incorporated the {\tt emcee} Python library \cite{foreman2013emcee}, providing a robust and parallelized implementation of Goodman and Weare’s affine invariant MCMC ensemble sampling algorithm \cite{GoodmanWeare}.

\section{Results}
\label{sec:results}
\begin{figure}[t]
    \centering
    \includegraphics{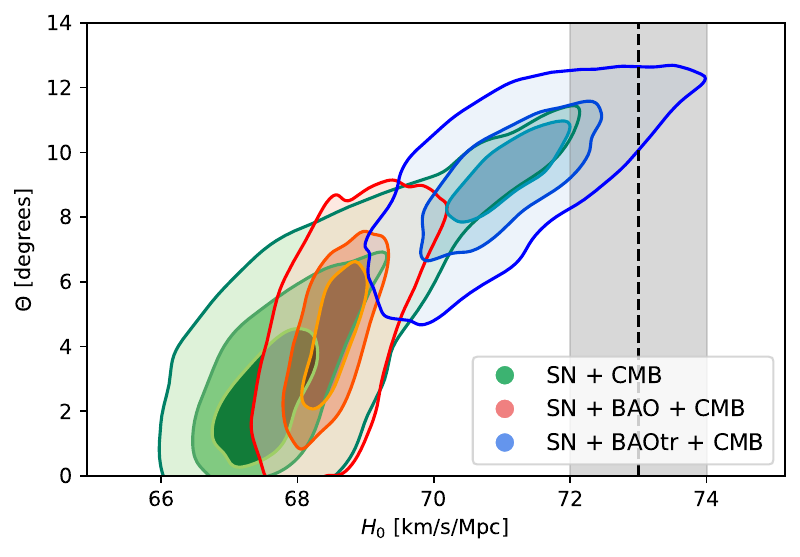}
    \caption{2D marginalized confidence level contours $(68 \, \% , 95 \, \% , 99 \, \%)$ in the $(H_0,\theta)$-plane for the general bimetric model.
    Gray: the SH0ES estimate of $H_0$. 
    Green: SN+CMB. With these data sets, the bimetric model yields a wide confidence contour, spanning from ``low'' values of $H_0$ to ``high'' values compatible with SH0ES.
    Blue: adding BAOtr. Notably, in this case the inferred value of $H_0$ is consistent with SH0ES. In other words, there is no Hubble tension in bimetric cosmology when BAOtr is used.
    Red: using ordinary 3D BAO data instead of BAOtr. In this case, $H_0$ centers around ``low'' values and a tension in $H_0$ is manifest.}
    \label{fig:main-plot}
\end{figure}

We present the results for the most general bimetric model. The free parameters and their flat prior ranges are presented in Tab.~\ref{tab:prior}.
In Fig.~\ref{fig:main-plot}, we show the 2D marginalized confidence level contours in the $(H_0 , \theta)$-plane for different combinations of data sets. The 2D marginalized confidence contours in the full parameter space is displayed in Fig.~\ref{fig:B01234}.
In Appendix~\ref{sec:submodels}, we also present the results for certain submodels.

\begin{table}[t]
 \centering
 \renewcommand{\arraystretch}{1.2}
 \begin{tabular}{r||ccccc}
 \hline\hline
 Model parameter: & $H_0$ & $\omega_b$ & $\theta$ & $\mfp$ & $\omegal$ \\
 \hline
 Prior: & $\mathrm{U}[50,85]$ & $\mathrm{U}(0,1]$ & $\mathrm{U}[0,\frac{\pi}{2}]$ & $\mathrm{U}(0,10^8]$ & $\mathrm{U}(0,1]$ \\
 \hline \hline
\end{tabular}
\caption{\label{tab:prior} Priors on the free model parameters for the most general ($B_0B_1B_2B_3B_4$) bimetric model. All priors are uniform, spanning the range indicated in the table. The upper limits of \(\alpha\) and \(\beta\) are set to 100 while their lower limits are effectively set by imposing the consistency constraints, explained in Section~\ref{sec:ConsConstr}.}
\end{table}

\begin{table}[t]
 \centering
 \renewcommand{\arraystretch}{1.2}
 \begin{tabular}{cl||ccccc}
 \hline\hline
 Model & CMB+SN+ & $H_0$ & $\theta$ (rad) & $\omegal$  & $\Delta$AIC & $\Delta$BIC\\
 \hline
 \lcdm & BAOtr & 68.9 $\pm$ 0.5    & 0 & 0.700 $\pm$ 0.006 & 0 & 0\\
 Bimetric &  BAOtr & 71.0 $\pm$ 0.9    & 0.16 $\pm$ 0.03 & 0.724 $\pm$ 0.010 & 1.1 & $-19.9$\\
 \hline
 \lcdm & BAOtr+SH0ES & 69.7 $\pm$ 0.5    & 0 & 0.709 $\pm$ 0.006 & 0 & 0\\
 Bimetric &  BAOtr+SH0ES & 71.9 $\pm$ 0.7    & 0.18 $\pm$ 0.02 & 0.732 $\pm$ 0.007 & 13.2 & $-7.8$\\
 \hline \hline
\end{tabular}
\caption{\label{tab:results} Best-fit values and $68 \, \%$ confidence errors for $H_0$, $\theta$, and $\omegal$. 
Recall that $H_0$ is given in units of km/s/Mpc.
With CMB+SN+BAOtr, the bimetric model exhibits a significant increase in $H_0$, thus being consistent with the local distance ladder measurement. 
\bic is the difference in the Bayesian information criterion between the \lcdm model and the bimetric model. A positive value of \bic indicates a favoured bimetric model. The corresponding also applies to \aic.}
\end{table}

In Fig.~\ref{fig:main-plot}, we see that SN+CMB yields a weakly constrained $H_0$, allowing both for ``high'' values compatible with SH0ES and for ``low'' values compatible with the standard \lcdm inverse distance ladder estimate. With SN+BAOtr+CMB, the inferred value $H_0 = 71.0 \pm 0.9$ is consistent with SH0ES.
This should be compared with $H_0 = 68.9 \pm 0.5$, which is the value obtained when a \lcdm model is fitted to the same data set.
In summary, utilizing transverse BAO data, the Hubble tension is alleviated in bimetric cosmology.

On the other hand, imposing ordinary 3D BAO data instead of BAOtr, there is only a slight increase in $H_0$ compared with \lcdm, so the tension remains, see Fig.~\ref{fig:main-plot}.
More specifically with SN+BAO+CMB, we get $H_0 = 68.6 \pm 0.5$ which is in the $4.4 \, \sigma$ tail of the SH0ES team estimate.

We conclude that there is a drastic difference in the inferred value of $H_0$ depending on whether BAOtr or 3D BAO is used. This indicates that the model dependence of the ordinary 3D BAO data can introduce a significant bias in $H_0$.

To understand why the bimetric model alleviates the tension when using BAOtr, we start by analyzing the \lcdm model. First, SNIa data constrain the shape of the expansion history at redshifts $z \leq 2.26$ while being agnostic to the absolute scale of the expansion rate ($H_0$).
Since the low-redshift expansion history is set by $\omegal$ in a \lcdm model, SNIa data constrain the value of $\omegal$. Further, the CMB angular scale $\theta_\mathrm{CMB}$ \eqref{eq:thetaCMB} is sensitive to the value of $H_0$. The observed value of $\theta_\mathrm{CMB}$ sets $H_0$ to be relatively low, that is, in tension with SH0ES. In other words, increasing $H_0$ to values compatible with SH0ES results in an increased angular scale $\theta_\mathrm{CMB}$, violating its observational constraints.

\begin{figure}[t]
    \centering
    \includegraphics[width=0.49\textwidth]{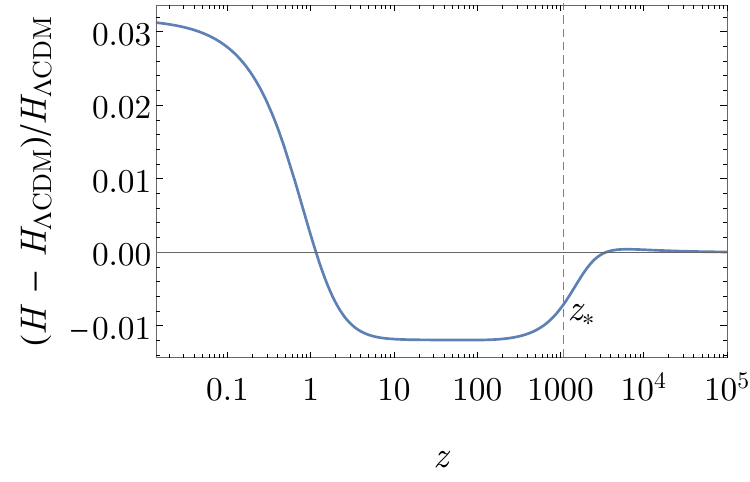}
    \includegraphics[width=0.49\textwidth]{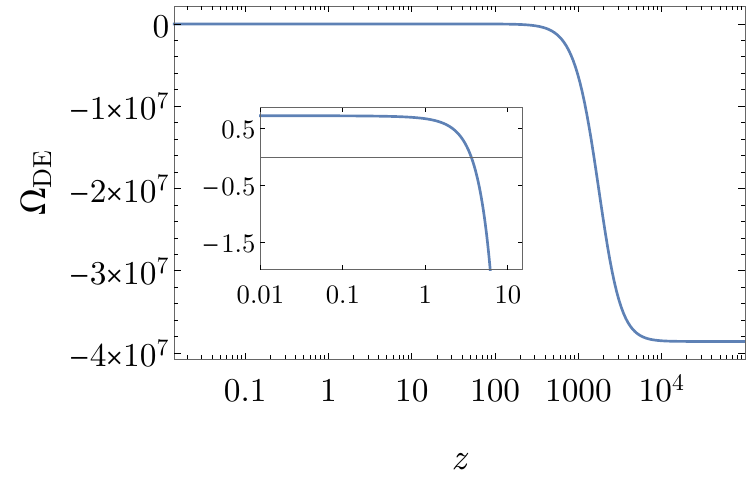}
    \caption{\emph{Left:} Relative difference in the expansion rate $H(z)$, comparing the CMB+SN+BAOtr best-fit bimetric and \lcdm models. The bimetric model exhibits $(H_0, \theta , \mfp , \omegal, \alpha ,\beta) = (71.0, 9.4^\circ, 5030, 0.724, 41, 46)$ and the \lcdm model has $(H_0, \omegal) = (68.9, 0.699)$. \emph{Right:} The dimensionless dark energy density $\omegade$ for the same bimetric model, defined as the physical energy density $\rho_\mathrm{DE}$ measured in units of the present-day critical energy density $\rho_c = 3 H_0^2/\kg$. As apparent, there is a transition from a negative cosmological constant at high redshifts to a positive cosmological constant at lower redshifts.}
    \label{fig:relH}
\end{figure}

\begin{figure}[t]
    \centering
    \includegraphics[width=0.49\textwidth]{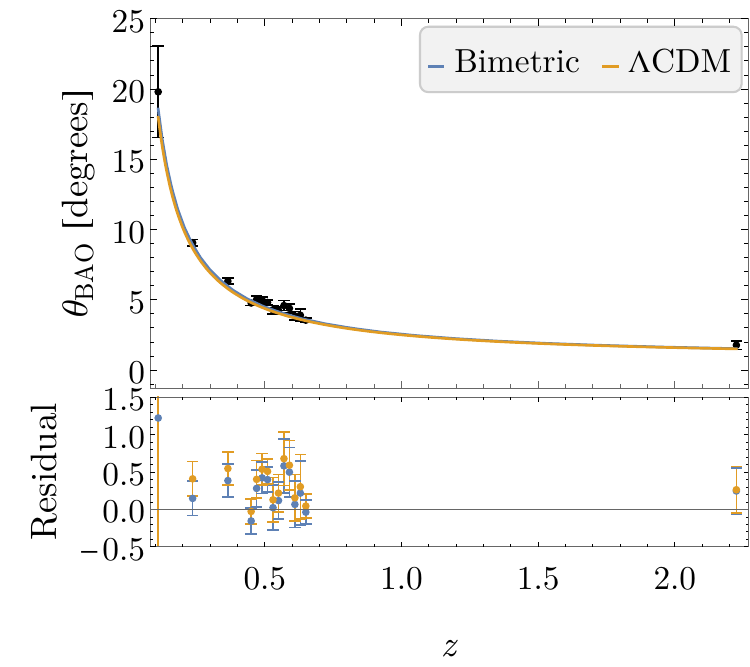}
    \includegraphics[width=0.49\textwidth]{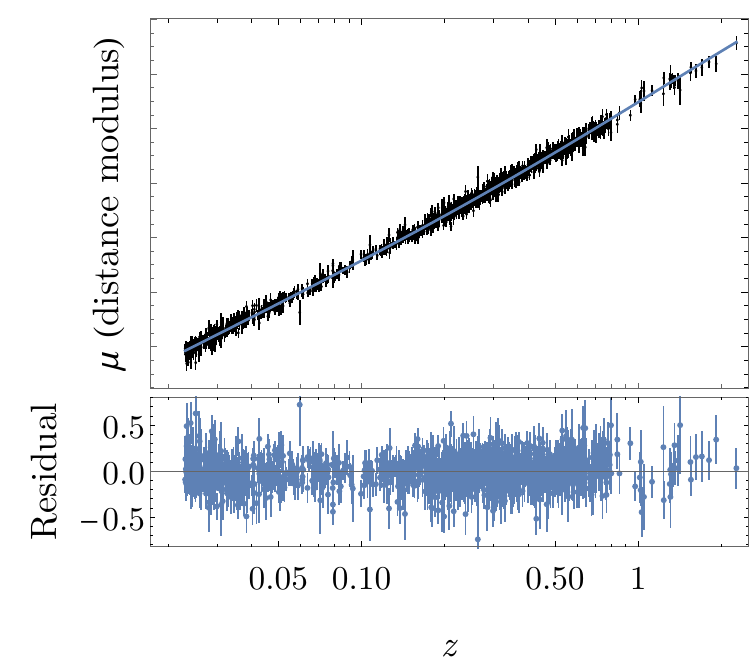}
    \caption{\emph{Left:} Angular BAO scale for a bimetric model and a \lcdm model. Data points are black. The bimetric model has the parameters $(H_0, \theta , \mfp , \omegal, \alpha ,\beta) = (71.0, 9.4^\circ, 5030, 0.724, 41, 46)$. The \lcdm model has $(H_0, \omegal) = (68.9, 0.699)$. By inspection, it can be seen that the \lcdm residuals have a systematic offset. This offset is reduced in the bimetric model.
    \emph{Right:} Distance modulus as a function of redshift for the same bimetric model. Since the SNIa data constrain only the shape of the distance modulus $\mu$ rather its intercept on the $y$-axis, we do not specify the absolute scale of the $y$-axis.}
    \label{fig:modelRes}
\end{figure}

However, this argument does not apply to bimetric cosmology due its increased flexibility. In this case, while the sound horizon stays the same, the expansion rate can be increased at small redshifts ($z \lesssim 1$) while decreased at intermediate redshifts ($1 \lesssim z \lesssim 1000$), compared with a \lcdm model. In this way, the angular scale $\theta_\mathrm{CMB}$ remains compatible with the observed value. An example is shown in Fig.~\ref{fig:relH} for the best-fit bimetric model.
However, it remains to explain why bimetric cosmology, in combination with BAOtr, actually prefers a cosmology with an increased expansion rate in the late universe. To understand this, in Fig.~\ref{fig:modelRes} we plot the angular BAO scale $\theta_\mathrm{BAO}$ for this model. Upon inspection of the residuals, it is evident that the \lcdm model prediction lies systematically below the observed values. This can be remedied by increasing the expansion rate at these redshifts. However, in \lcdm this is not possible due to the tight constraints on $H_0$ from the CMB. In bimetric cosmology on the other hand, this is possible, as explained above.
In short, BAOtr prefers an increased expansion rate in the late universe which can be accommodated in bimetric cosmology as opposed to \lcdm. This is possible without spoiling CMB data due to a decrease in the expansion rate at intermediate redshifts.

With 3D BAO data on the other hand, there is no such offset between the \lcdm model prediction and data. So, 3D BAO does not prefer an increased expansion rate at low redshifts, explaining the difference in $H_0$ between BAOtr and 3D BAO. In summary, the choice of BAO data set (BAOtr or 3D BAO) effectively splits the SN+CMB confidence contour in Fig.~\ref{fig:main-plot} into two pieces with the model independent BAOtr piece being compatible with the SH0ES measurement and the 3D BAO piece assuming ``low'' $H_0$ values in tension with SH0ES. 
The split between the 2D BAO and 3D BAO contours can be explained by the slight tension between these two data sets, which has been discussed and quantified for example in Refs.~\cite{Camarena:2019rmj,Gomez-Valent:2023uof,Favale:2024sdq}.

Finally, we check that the shape of the SH0ES-compatible bimetric expansion history is compatible with the SNIa data. To demonstrate this, in Fig.~\ref{fig:modelRes} we plot the distance modulus, defined as the difference between the apparent magnitude and the absolute magnitude, $\mu = m - M$. The theory prediction for this quantity is given by
\begin{equation}
    \mu(z) = 5 \log_{10} \mathcal{D}_L(z) - 5 \log_{10} H_0 + 25.
\end{equation}
As shown in Fig.~\ref{fig:modelRes}, the bimetric model exhibits an expansion rate which is compatible also with SNIa data.

To quantify the degree of success of the bimetric model compared with \lcdm, we assess the goodness-of-fit versus the number of model parameters. Two common measures of this are the Akaike information criterion (AIC) and the Bayesian information criterion (BIC) \cite{Akaike:1974vps,Schwarz:1978tpv}.
The AIC is defined as 
\begin{equation}
    \label{eq:AIC}
    \mathrm{AIC} = 2 N_\mathrm{param} + \chi^2_\mathrm{min}.
\end{equation}
The BIC approximates the Bayesian evidence and is defined as
\begin{equation}
    \label{eq:BIC}
    \mathrm{BIC} = N_\mathrm{param} \ln N_\mathrm{data} + \chi^2_\mathrm{min},
\end{equation}
where $N_\mathrm{param}$ is the number of model parameters and $N_\mathrm{data}$ is the number of data points.
Thus, the AIC and the BIC quantify how well a model is performing by balancing the goodness-of-fit of the model against its complexity, penalizing complex models that exhibit overfitting. In fact, the BIC exhibits a stronger penalty on models with more parameters compared with the AIC.

The degree to which one model is preferred over another is quantified by
\begin{subequations}
    \begin{align}
        \Delta \mathrm{AIC} &= \mathrm{AIC}_{\Lambda \mathrm{CDM}}-\mathrm{AIC}_\mathrm{bimetric}, \\
        \Delta \mathrm{BIC} &= \mathrm{BIC}_{\Lambda \mathrm{CDM}}-\mathrm{BIC}_\mathrm{bimetric},
    \end{align}
\end{subequations}
so a positive value of \aic or \bic favours bimetric cosmology over \lcdm, and vice versa. Following Jeffrey--Raftery's terminology, \bic in the range 0--2 is referred to as ``weak evidence'', 2--6 as ``positive evidence'', 6--10 as ``strong evidence'', and $>10$ as ``very strong evidence'' \cite{raftery}.

As seen in Tab.~\ref{tab:results}, we are facing the ambiguous situation where the bimetric model is preferred by the AIC but the \lcdm model is preferred by the BIC. If we focus on the inverse distance ladder (CMB+SN+BAOtr, i.e. without the SH0ES prior), there is a slight preference for the bimetric model according to the AIC whereas there is a strong preference for the \lcdm model according to the BIC.
Taken together, the two information criteria are indecisive whereas to which model that exhibits the best performance. 

This demonstrates the limitations of the information criteria in distinguishing between models. 
However, it should be stressed that the bimetric model has the advantage of the inverse distance ladder predicting a Hubble constant which is consistent with the local distance ladder measurement.

As seen in Fig.~\ref{fig:relH}, the bimetric model alleviates the Hubble tension by increasing the expansion rate at low redshifts ($z \lesssim 1$), compensating for this by a decreased expansion rate at intermediate redshifts ($1 \lesssim z \lesssim 1000$) so that the CMB angular scale remains invariant.
This expansion history is realized by the bimetric dark energy density $\omegade$ exhibiting a negative cosmological constant at high redshifts with a transition to a positive cosmological constant at low redshifts.
In the example in Fig.~\ref{fig:relH}, $\omegade$ consititutes a few percent of the total energy budget at redshifts $z \gtrsim 2$, to then drop below the subpercent level at redshifts $z \gtrsim 1500$. At low redshifts ($z \lesssim 1$), $\omegade$ contributes significantly to the expansion history.

As such, with $\omegade$ increasing with time, it constitutes a phantom dark energy.
Since $\omegade$ mimics a cosmological constant also in the early universe, the bimetric modification to the expansion history can be ignored in the early universe. This can be seen in Fig.~\ref{fig:relH} as the expansion rate differs $< 1 \, \%$ at prerecombination redshifts $z > z_*$. This also means that the sound horizon at recombination and at the baryon drag epoch are the same as in the \lcdm model.\footnote{For the best-fit bimetric cosmology, shown in Fig.~\ref{fig:relH}, the sound horizons are $r_s(z_*) = 144.4 \, \mathrm{Mpc}$ and $r_s(z_d) = 147.0 \, \mathrm{Mpc}$.}
Accordingly, with $\theta_\mathrm{CMB}$ fixed from observations, the distance to the last scattering surface $D_A(z_*)$ also remains invariant.

\section{Discussion}
Apparently, the choice between 2D BAO and 3D BAO makes a decisive difference with respect to the inferred value of $H_0$ with the more conservative option (BAOtr) alleviating the tension. This calls for further analysis of a possible bias in the 3D BAO data.
Ideally the 3D BAO data reduction should be redone assuming a bimetric cosmology. The goal of such an analysis would be to decrease the relatively large $\sim 10 \, \%$ errors in BAOtr to $\sim 1 \, \%$ errors as in the ordinary 3D BAO data, to see whether any Hubble tension-solving cosmologies remain viable.
In Ref.~\cite{Sanz-Wuhl:2024uvi}, such a re-analysis was carried out in the case of non-flat cosmology with the result that the 3D BAO data is robust with respect to the choice of spatial curvature.
In Ref.~\cite{Carter:2019ulk}, the authors analyzed the effects of assuming an inadequate fiducial cosmology for a set of $w$CDM cosmologies. It was concluded that it has little impact on the BAO shift parameters but leads to a notable misjudgement of the measured errors.
With bimetric cosmology providing a solution to the Hubble tension with 2D BAO, we call for such a re-analysis of the 3D BAO data for this set of cosmologies. This will likely be demanding. Nevertheless, it is necessary to fully establish bimetric cosmology as an explanation of the Hubble tension. 

With BAOtr, the Hubble constant inferred from the inverse distance ladder is brought into $2 \, \sigma$ agreement with the Cepheid-based local distance ladder estimate from the SH0ES team. The residual $2 \, \sigma$ difference might be explained due to the local effects of bimetric gravity, effectively providing a fifth force that can recalibrate the local distance ladder along the lines of Refs.~\cite{Desmond:2019ygn,Hogas:2023pjz}. In other words, bimetric gravity may not only re-calibrate the inverse distance ladder but also the local distance ladder. This should be assessed in a separate work.

The effectiveness of bimetric cosmology in alleviating the Hubble tension should also not be viewed in isolation. To be consistent, the theory parameters singled out by the inverse distance ladder should describe a gravitational phenomenology in accordance with complementary tests of gravity.
In Fig.~8.1 of Ref.~\cite{Hogas:2022owf}, observational constraints from the following sources are compiled: solar-system tests, strong gravitational lensing (SGL) by galaxies, gravitational waves, and the abundance of the light elements (Big Bang nucleosynthesis).

With these additional probes, there remain viable regions in the parameter space in the range $1 \lesssim \mfp \lesssim 10^4$.\footnote{There are also windows of viable Fierz--Pauli masses in the ranges $10^6 \lesssim \mfp \lesssim 10^8$ and $\mfp \gtrsim 10^{32}$.} Interestingly, the allowed region overlaps with the region where the Hubble tension is alleviated.
The lower limit $\mfp \gtrsim 1$ is due to the Higuchi bound which is a theoretical requirement ensuring the absence of the Higuchi ghost \cite{Higuchi:1986py}.
The upper limit $\mfp \lesssim 10^4$ is due to SGL. Recall that in the MCMC method, we set the likelihood to zero whenever there is no Vainshtein screening mechanism. In other words, the existence of a Vainshtein screening mechanism is guaranteed, making the region $1 \lesssim \mfp \lesssim 10^4$ observationally viable. 

As an example, with $\mfp \sim 1000$, the Vainshtein radius is $r_V \sim 10 \, \mathrm{kpc} $ for a $M = 10^{12} M_\odot$ galaxy, meaning that the typical lensing radius is of similar magnitude as the Vainshtein screening radius. This implies that $\mfp \sim 1000$ is on the boundary of what is observational allowed by SGL. For a detailed analysis, see Ref.~\cite{Guerrini:2023pre}. With future cosmological surveys such as Euclid, the sample size of strong gravitational lenses will increase by several orders of magnitude.
This will probe larger portions of the parameter space (lower values of $\mfp$) where the Hubble tension-solving models are located.

Another, less significant, tension that has gained interest within the last decade is the tension between the observed distribution of galaxies predicted by CMB observations, assuming a \lcdm cosmology, and for example large-scale structure surveys and weak lensing. This is quantified by the parameter $S_8 = \sigma_8 \sqrt{\Omega_{m,0}/0.3}$ where $\sigma_8$ is the present-day average amplitude of matter fluctuations at $8h^{-1} \, \mathrm{Mpc}$.
In bimetric cosmology, such as the model $(H_0, \theta , \mfp , \omegal, \alpha ,\beta) = (71.0, 9.4^\circ, 5030, 0.724, 41, 46)$ studied in Section~\ref{sec:results}, the present-day matter density $\Omega_{m,0}$ is greater than in a \lcdm model. For this model, $\Omega_{m,0} = 0.284$. The decrease in $\Omega_{m,0}$ is due to the fact that $\omegal$ is the effective cosmological constant that $\omegade$ approaches at future infinity, and thus assumes a greater value than the present-day value $\Omega_\mathrm{DE,0}$. With a decreased present-day matter density, $S_8$ decreases, also easing the $S_8$ tension, assuming that $\sigma_8$ is constant. 
However, one should be very careful to draw such a conclusion at this stage since there is currently no framework that allows to predict structure formation in this theory \cite{Konnig:2015lfa,Aoki:2015xqa,Mortsell:2015exa,Akrami:2015qga,Hogas:2019ywm,Luben:2019yyx}. So, at present we cannot justify the assumption of $\sigma_8$ being unchanged in this theory. 

We have focused on the most general bimetric model. In Tab.~\ref{tab:resultsFull} and Tab.~\ref{tab:resultsSH0ESFull} in Appendix~\ref{sec:submodels}, we present the corresponding results for all viable submodel, that is, $B_0 B_1 B_2 B_3$, $B_1 B_2 B_3 B_4$, and $B_1 B_2 B_3$. The four-parameter model $B_0 B_1 B_2 B_3$ is performing better than $B_1 B_2 B_3 B_4$ with respect to the Hubble tension. Due to the decreased number of theory parameters, these models are slightly favoured over the general model when assessed by the BIC. The most restricted submodel ($B_1 B_2 B_3$) on the other hand yields a Hubble constant which is in tension with SH0ES, just as the \lcdm model. 

In the literature, there are alternative cosmological models exhibiting qualitatively similar expansion histories as bimetric cosmology. Accordingly, the conclusions in the current paper applies, with appropriate adaptations, also to these models. Some of them are discussed below.

\paragraph{Minimal theories of bigravity (MTBG).} This set of theories \cite{DeFelice:2020ecp} share the cosmological background solutions with bimetric gravity. Therefore, the results in the current paper apply straightforwardly also to MTBG.

\paragraph{$\Lambda_s$CDM.} This is a phenomenological model exhibiting a sudden transition between a negative and a positive cosmological constant at some redshift \cite{Akarsu:2021fol,Akarsu:2022typ,Akarsu:2023mfb}.\footnote{The influence of a more general framework, dubbed $\Lambda$XCDM, on the Hubble tension was studied in Ref.~\cite{Gomez-Valent:2024tdb}.} Thus, bimetric gravity accomodates this set of models in a theoretically robust framework, only with a smooth transition instead of a discrete one. Similar to the results in the present paper, it was shown that the $\Lambda_s$CDM model yields a value for $H_0$ which is compatible with SH0ES when using BAOtr data, thereby alleviating the tension. 
The main difference between bimetric cosmology and $\Lambda_s$CDM is that the latter introduces only one additional model parameter compared with the four in bimetric cosmology. This results in an unambiguous preference for $\Lambda_s$CDM, also by the BIC.
A generalized version of the $\Lambda_s$CDM model, dubbed $\Lambda$XCDM, was recently studied in Ref.~\cite{Gomez-Valent:2024tdb}. The authors show that the $\Lambda$XCDM is even outperforming the successful $\Lambda_s$CDM model, primarily by increasing the quality of fit to SNIa data. We note however that bimetric gravity does not accommodate the $\Lambda$XCDM model, due to the latter exhibiting a quintessence-like dark energy in the late universe. This is a feature that cannot be achieved within bimetric cosmology alone.

\paragraph{Quadratic bimetric gravity.} A ghost-free generalization of bimetric gravity was studied in Refs.~\cite{Gialamas:2023aim,Gialamas:2023fly,Gialamas:2023lxj}. Here, in addition to the two Einstein--Hilbert terms of each metric, one adds, for each metric, a term which is quadratic in the Ricci scalar, similar to a Starobinsky model \cite{Starobinsky:1980te}. The interaction term remains the same as in the standard Hassan--Rosen bimetric action \eqref{HR_action}. Accordingly, the set of bimetric cosmologies studied in the current work is a subset of the expansion histories provided by the theory of quadratic bimetric gravity, obtained by setting the coefficients of the quadratic terms to zero. Therefore, the alternative expansion histories featured in the current paper are also to be found in quadratic bimetric gravity.

\paragraph{Phenomenological emergent dark energy.} Ref.~\cite{Li:2019yem} proposed a phenomenological dark energy model where the dark energy vanishes in the early universe. In the late universe, the energy density grows, becoming influential at redshifts $z \lesssim 1$ to then approach a cosmological constant equation of state towards future infinity. In Refs.~\cite{Li:2020ybr,Yang:2021eud,Hernandez-Almada:2024ost}, the influence of this model on the Hubble tension was investigated. Interestingly, this phenomenological model is very similar to the self-accelerating bimetric cosmologies (i.e., $B_0=0$), where $\omegade \to 0$ as $z \to \infty$ but approaching a cosmological constant $\omegade \to \omegal$ towards future infinity \cite{Hogas:2021fmr,Hogas:2021lns}. In other words, bimetric gravity accommodates this phenomenological model in a robust theoretical framework.

\section{Conclusions}
Recent works in the literature indicate that the Hubble tension may be alleviated if standard 3D BAO data is replaced by of 2D BAO data \cite{Li:2020ybr,Yang:2021eud,Akarsu:2021fol,Akarsu:2022typ,Akarsu:2023mfb,Gomez-Valent:2024tdb,Hernandez-Almada:2024ost}. Moreover, 2D BAO has the advantage of being only weakly model-dependent in the data reduction, thus offering a set of data points that can coherently be used to probe expansion histories beyond \lcdm \cite{Sanchez:2010zg,Carvalho:2015ica,Alcaniz:2016ryy,deCarvalho:2017xye,Carvalho:2017tuu,Nunes:2020hzy,deCarvalho:2021azj}.
The effects of 2D BAO data on the Hubble tension has been studied in the case of phenomenological cosmologies such as the $\Lambda_s$CDM model \cite{Akarsu:2021fol,Akarsu:2022typ,Akarsu:2023mfb} and phenomenological emergent dark energy (PEDE) \cite{Li:2019yem}. Both of these models exhibit an increased expansion rate relative to \lcdm at redshifts $z \lesssim 1$, thereby easing the Hubble tension. 

In the present work, we study bimetric gravity which is a consistent theory of gravity that accommodates the $\Lambda_s$CDM model as well as PEDE, in addition to a much wider range of expansion histories.
We show that, in bimetric cosmology, the inverse distance ladder with SN+2D BAO+CMB yields $H_0 = (71.0 \pm 0.9) \, \mathrm{km/s/Mpc}$ which is compatible with SH0ES $H_0 = (73.0 \pm 1.0) \, \mathrm{km/s/Mpc}$ at the $2 \, \sigma$ border, thus alleviating the Hubble tension. 
On the other hand, with ordinary 3D BAO data bimetric cosmology eases the tension only very slightly, with a $4.4 \, \sigma$ tension remaining.

Upon evaluating the Akaike Information Criterion (AIC) and the Bayesian Information Criterion (BIC), we encounter an ambiguous result: the AIC favors the bimetric model, while the BIC favors the \lcdm model. This discrepancy highlights the limitations of both criteria in definitively determining which model performs better. Nonetheless, it is worth emphasizing that the bimetric model offers a key advantage: the inverse distance ladder (with 2D BAO) predicts a Hubble constant consistent with measurements from the local distance ladder.

A scientific advantage with bimetric cosmology, as opposed to the phenomenological models, is that the cosmological analysis can be supplemented with complementary observational tests of gravity to ensure that it offers a coherent picture of gravity at all observable scales. This includes solar-system tests, gravitational waves, strong gravitational lensing by galaxies, and the abundance of the light elements (Big Bang nucleosynthesis). We show that a Hubble-tension solving cosmology can indeed fit coherently within a gravitational theory.
Unfortunately, there is currently no framework for calculating the growth of structure in this theory. So, at present the effect of this theory on the $S_8$ tension is unclear although a very preliminary assessment shows promising results also for this (less major) tension.

We have shown that, in bimetric cosmology, the inverse distance ladder yields significantly different values for the Hubble constant depending on whether 3D BAO data or 2D BAO data is used. This is due to 2D BAO preferring a greater angular BAO scale than 3D BAO. Accordingly, 2D BAO prefers a higher value for $H_0$. In bimetric cosmology, this can be achieved by increasing the expansion rate (relative to \lcdm) at redshifts $z \lesssim 1$, compensating for this by decreasing the expansion rate at redshifts $z \gtrsim 1$ so as to maintain the well-constrained angular CMB scale. (The sound horizon remains unchanged.) This expansion history is realized by a smooth transition from a negative cosmological constant at $z \gtrsim 1$ to a positive one at $z \lesssim 1$.

A possible explanation for the difference in $H_0$ between 2D BAO and 3D BAO is that the latter might bias $H_0$ low due to the assumption of \lcdm cosmology during the 3D BAO data reduction. This is apparently an assumption which is incompatible with the alternative expansion histories studied here. If so, this could explain of the Hubble tension. To definitely establish this as an explanation, we call for a reanalysis of the 3D BAO data reduction under less restrictive model assumptions.

\appendix
\section{Ratio of the scale factors: equation of motion}
\label{sec:MixedEqs}
The quartic polynomial governing the evolution of $y$, the ratio of the scale factors, is given by,
\begin{align}
\label{eq:yPoly}
	& - \frac{1}{3} \cos^2 \theta \, \mfp^2 (1+2\alpha+\beta) \nonumber \\
	&+ \left[\Omega_\mathrm{tot}(z) + \omegal + \mfp^2 \left(\cos^2 \theta \, (\alpha + \beta) - \sin^2 \theta \, \left(1+\alpha + \frac{\beta}{3}\right)\right)\right] y(z) \nonumber \\ 
	& + \mfp^2 \left[-\cos^2 \theta \, \beta + \sin^2 \theta \, (1+2\alpha+\beta)\right] y^2(z) \nonumber\\
	&-\left[\omegal  + \frac{1}{3} \mfp^2 \left(\cos^2 \theta \, (-1+\alpha-\beta) + 3\sin^2 \theta \, (\alpha+\beta)\right)\right] y^3(z) \nonumber \\
	&+ \frac{1}{3} \sin^2 \theta \, \mfp^2 \beta \, y^4(z) = 0,
\end{align}
where $\Omega_\mathrm{tot}(z) \equiv \Omega_m(z) + \Omega_r(z)$. In the data analysis, we solve this equation numerically for each redshift, choosing the finite branch solution. That is, the solution satisfying $0 \leq y \leq 1$.

Evaluating eq.~\eqref{eq:yPoly} and eq.~\eqref{eq:Friedmann} at $z=0$ (present day), the equations can be combined to yield an equation involving only $y_0$ and the physical parameters. Here, $y_0$ is the value of $y$ at $z=0$. The quartic term cancels and the equation reads,
\begin{align}
\label{eq:ycub}
	& \frac{1}{3} \cos^2 \theta \, \mfp^2 (1+2\alpha+\beta) - \left[1 + \cos^2 \theta \, \mfp^2 \, (\alpha + \beta) \right]y_0 + \cos^2 \theta \, \mfp^2 \, \beta \, y_0^2 \nonumber\\
	+ &\left[\omegal  + \frac{1}{3} \cos^2 \theta \,  \mfp^2 (-1+\alpha-\beta)\right] y_0^3 =0.
\end{align}

\section{Complementary results}
\label{sec:submodels}
In Tab.~\ref{tab:resultsFull} and Tab.~\ref{tab:resultsSH0ESFull} we display results complementary to Tab.~\ref{tab:results}, including the viable bimetric submodels $B_0 B_1 B_2 B_3$, $B_1 B_2 B_3 B_4$, and $B_1 B_2 B_3$.

\begin{table}[H]
 \centering
 \renewcommand{\arraystretch}{1.2}
 \begin{tabular}{cl||cccccc}
 \hline\hline
 Model & CMB+SN+ & $H_0$ & $\theta$ (rad) & $\omegal$ & $\mfp$ & $\Delta$AIC  & $\Delta$BIC\\
 \hline
 \multirow{4}{*}{\lcdm} & \multirow{2}{*}{BAO} & 68.3 & \multirow{2}{*}{0}  & 0.691 & \multirow{2}{*}{---} & \multirow{2}{*}{0} & \multirow{2}{*}{0}\\
 & & $\pm 0.4$ & & $\pm 0.005$ & &\\
    & \multirow{2}{*}{BAOtr} & 68.9 & \multirow{2}{*}{0} & 0.700 & \multirow{2}{*}{---} & \multirow{2}{*}{0} & \multirow{2}{*}{0}\\
 & & $\pm 0.5$ & & $\pm 0.006$ & &\\
 \hline
 \multirow{4}{*}{$B_0B_1B_2B_3B_4$} & \multirow{2}{*}{BAO} & 68.6 & 0.078 & 0.695 & \multirow{2}{*}{$< 1.9 \times 10^4$} & \multirow{2}{*}{$-7.1$} & \multirow{2}{*}{$-28.0$}\\
 & & $\pm 0.5$ & $\pm 0.036$ & $\pm 0.006$ & &\\
 & \multirow{2}{*}{BAOtr} & 71.0 & 0.163 & 0.724 & \multirow{2}{*}{$< 1.7 \times 10^4$} &  \multirow{2}{*}{$1.1$} & \multirow{2}{*}{$-19.9$}\\
 & & $\pm 0.9$ & $\pm 0.028$ & $\pm 0.010$ & & \\
 \hline
  \multirow{4}{*}{$B_0B_1B_2B_3$} & \multirow{2}{*}{BAO} & 68.5 & 0.088 & 0.695 & \multirow{2}{*}{$< 8.0 \times 10^4$} & \multirow{2}{*}{$-3.8$} & \multirow{2}{*}{$-20.6$}\\
  & & $\pm 0.6$ & $\pm 0.040$ & $\pm 0.007$ & & \\
  & \multirow{2}{*}{BAOtr} & 70.9 & 0.171 & 0.726 & \multirow{2}{*}{$< 1.6 \times 10^4$} & \multirow{2}{*}{$3.0$} & \multirow{2}{*}{$-12.7$}\\
  & & $\pm 1.0$ & $\pm 0.059$ & $\pm 0.015$ & & \\
  \hline
 \multirow{4}{*}{$B_1B_2B_3B_4$} & \multirow{2}{*}{BAO} & 68.3 & 0.015 & 0.692 & \multirow{2}{*}{---} & \multirow{2}{*}{$-8.6$} & \multirow{2}{*}{$-21.3$} \\
 & & $\pm 0.4$ & $\pm 0.035$ & $\pm 0.005$ & & \\
 & \multirow{2}{*}{BAOtr} & 69.3 & 0.100 & 0.707 & \multirow{2}{*}{---} & \multirow{2}{*}{$-2.6$} & \multirow{2}{*}{$-18.3$}\\
 & & $\pm 0.8$ & $\pm 0.073$ & $\pm 0.013$ & & \\
 \hline
\multirow{4}{*}{$B_1B_2B_3$} & \multirow{2}{*}{BAO} & 68.2 & \multirow{2}{*}{$< 10^{-2.6}$} & 0.691 & \multirow{2}{*}{---} & \multirow{2}{*}{$-4.0$} & \multirow{2}{*}{$-14.5$} \\
& & $\pm 0.4$ & & $\pm 0.005$ & & \\
& \multirow{2}{*}{BAOtr} & 68.9 & \multirow{2}{*}{$< 10^{-2.6}$} & 0.699 & \multirow{2}{*}{---} & \multirow{2}{*}{$-2.9$} & \multirow{2}{*}{$-13.4$} \\
& & $\pm 0.5$ & & $\pm 0.006$ & & \\
 \hline \hline
\end{tabular}
\caption{\label{tab:resultsFull} Best-fit values and $68 \, \%$ confidence errors for $H_0$, $\theta$, and $\omegal$ due to CMB+SN+BAO(tr) data. The remaining theory parameters, $\alpha$ and $\beta$ are unconstrained as seen in Figs.~\ref{fig:B01234}--\ref{fig:B123}, thus not tabulated. In the case where $\theta = 0$ is contained within the credible interval, $\mfp$ is unconstrained.}
\end{table}

\begin{table}[H]
 \centering
 \renewcommand{\arraystretch}{1.2}
 \begin{tabular}{cl||cccccc}
 \hline\hline
 Model & CMB+SN+ & $H_0$ & $\theta$ (rad) & $\omegal$ & $\mfp$ & $\Delta$AIC  & $\Delta$BIC\\
 \hline
 \multirow{4}{*}{\lcdm} & \multirow{2}{*}{BAO+SH0ES} & 68.9 & \multirow{2}{*}{0} & 0.698 & \multirow{2}{*}{---} & \multirow{2}{*}{0} & \multirow{2}{*}{0} \\
 & & $\pm 0.3$ & & $\pm 0.004$ & & \\
    & \multirow{2}{*}{BAOtr+SH0ES} & 69.7 & \multirow{2}{*}{0} & 0.709 & \multirow{2}{*}{---} & \multirow{2}{*}{0} & \multirow{2}{*}{0} \\
    & & $\pm 0.5$ & & $\pm 0.006$ & & \\
 \hline
 \multirow{4}{*}{$B_0B_1B_2B_3B_4$} & \multirow{2}{*}{BAO+SH0ES} & 69.4 & 0.119 & 0.704 & \multirow{2}{*}{$< 9.9 \times 10^4$} & \multirow{2}{*}{$-1.8$} & \multirow{2}{*}{$-22.7$} \\
 & & $\pm 0.5$ & $\pm 0.031$ & $\pm 0.006$ & & \\
 & \multirow{2}{*}{BAOtr+SH0ES} & 71.9 & 0.178 & 0.732 & \multirow{2}{*}{$< 8.3 \times 10^3$} & \multirow{2}{*}{$13.2$} & \multirow{2}{*}{$-7.8$} \\
 & & $\pm 0.7$ & $\pm 0.023$ & $\pm 0.007$ & & \\
 \hline
  \multirow{4}{*}{$B_0B_1B_2B_3$} & \multirow{2}{*}{BAO+SH0ES} & 69.2 & 0.109 & 0.704 & \multirow{2}{*}{$< 1.4 \times 10^5$} & \multirow{2}{*}{0.2} & \multirow{2}{*}{$-15.5$} \\
  & & $\pm 0.6$ & $\pm 0.038$ & $\pm 0.006$ & & \\
  & \multirow{2}{*}{BAOtr+SH0ES} & 71.8 & 0.183 & 0.733 & 
\multirow{2}{*}{$< 1.1 \times 10^4$} & \multirow{2}{*}{15.0} & \multirow{2}{*}{$-0.7$} \\
  & & $\pm 0.8$ & $\pm 0.022$ & $\pm 0.008$ & & \\
  \hline
 \multirow{4}{*}{$B_1B_2B_3B_4$}  &\multirow{2}{*}{BAO+SH0ES} & 69.0 & 0.048 & 0.701 & \multirow{2}{*}{---} & \multirow{2}{*}{$-1.5$} & \multirow{2}{*}{$-17.2$} \\
 & & $\pm 0.4$ & $\pm 0.057$ & $\pm 0.006$ & & \\
 & \multirow{2}{*}{BAOtr+SH0ES} & 70.6 & 0.173 & 0.723 & \multirow{2}{*}{---} & \multirow{2}{*}{6.7} & \multirow{2}{*}{$-9.0$} \\
 & & $\pm 0.7$ & $\pm 0.067$ & $\pm 0.010$ & & \\
 \hline
\multirow{4}{*}{$B_1B_2B_3$} & \multirow{2}{*}{BAO+SH0ES} & 68.8 & \multirow{2}{*}{$< 10^{-2.5}$} & 0.698 & \multirow{2}{*}{---} & \multirow{2}{*}{$-2.6$} & \multirow{2}{*}{$-13.1$} \\
& & $\pm 0.4$ & & $\pm 0.005$ & & \\
& \multirow{2}{*}{BAOtr+SH0ES} & 69.7 & \multirow{2}{*}{$< 10^{-2.5}$} & 0.709 & \multirow{2}{*}{---} & \multirow{2}{*}{$-1.0$} & \multirow{2}{*}{$-11.5$} \\
& & $\pm 0.5$ & & $\pm 0.006$ & & \\
 \hline \hline
\end{tabular}
\caption{\label{tab:resultsSH0ESFull} Best-fit values and $68 \, \%$ confidence errors for $H_0$, $\theta$, and $\omegal$ due to CMB+SN+BAO(tr)+SH0ES. That is, including the $H_0$ prior from SH0ES. The remaining theory parameters, $\alpha$ and $\beta$ are unconstrained as seen in Figs.~\ref{fig:B01234}--\ref{fig:B123}, thus not tabulated. In the case where $\theta = 0$ is contained within the credible interval, $\mfp$ is unconstrained.}
\end{table}

\begin{figure}[H]
    \centering
    \includegraphics[width=1\textwidth]{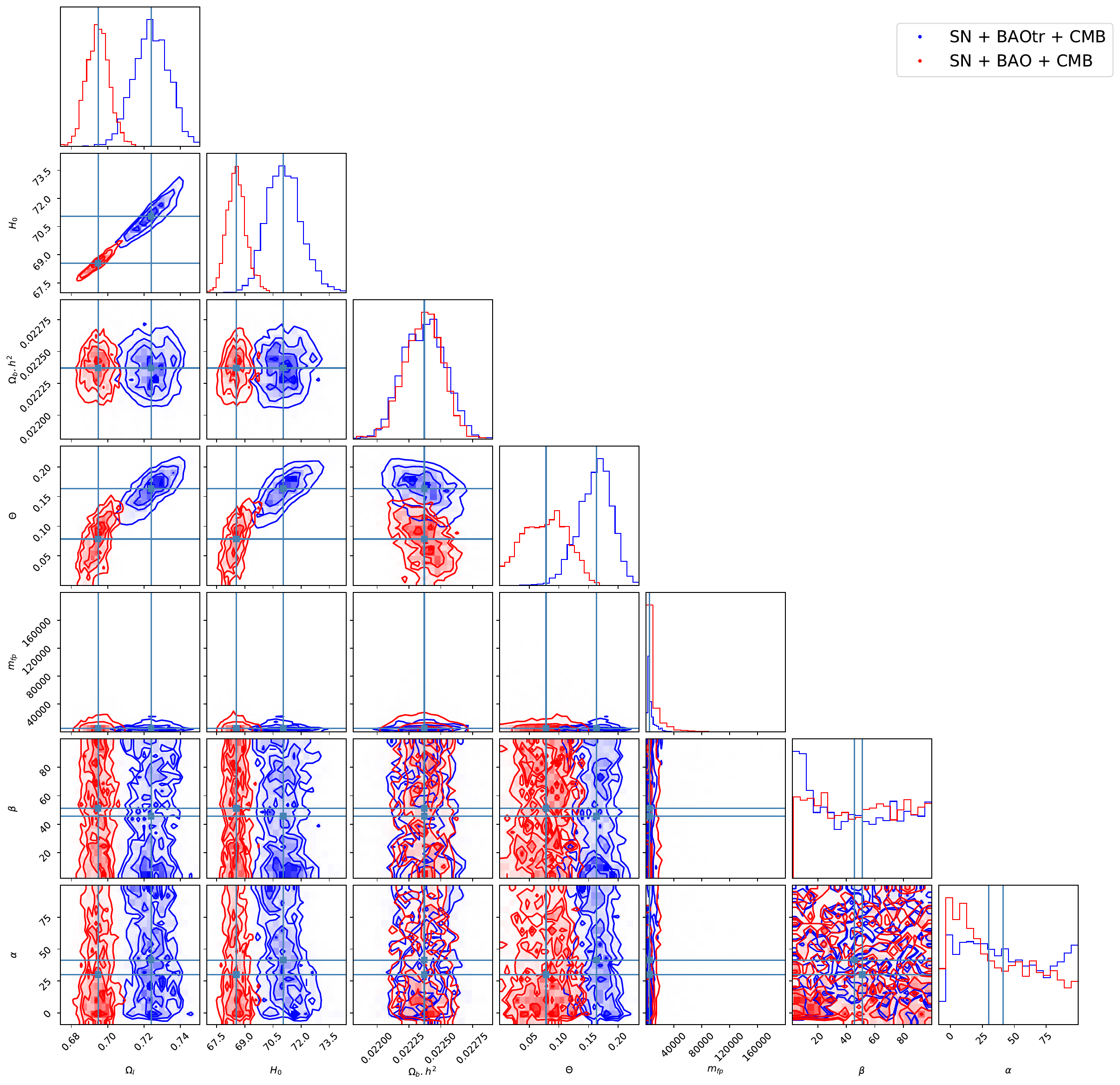}
    \caption{The corner plot shows the 2D marginalized CL contours for the most general bimetric model ($B_0B_1B_2B_3B_4$). It is evident that the inclusion of BAOtr instead of ordinary 3D BAO results in a noticeable increment of the estimated value of the Hubble constant $H_0$.}
    \label{fig:B01234}
\end{figure}

\begin{figure}[H]
    \centering
    \includegraphics[width=1\textwidth]{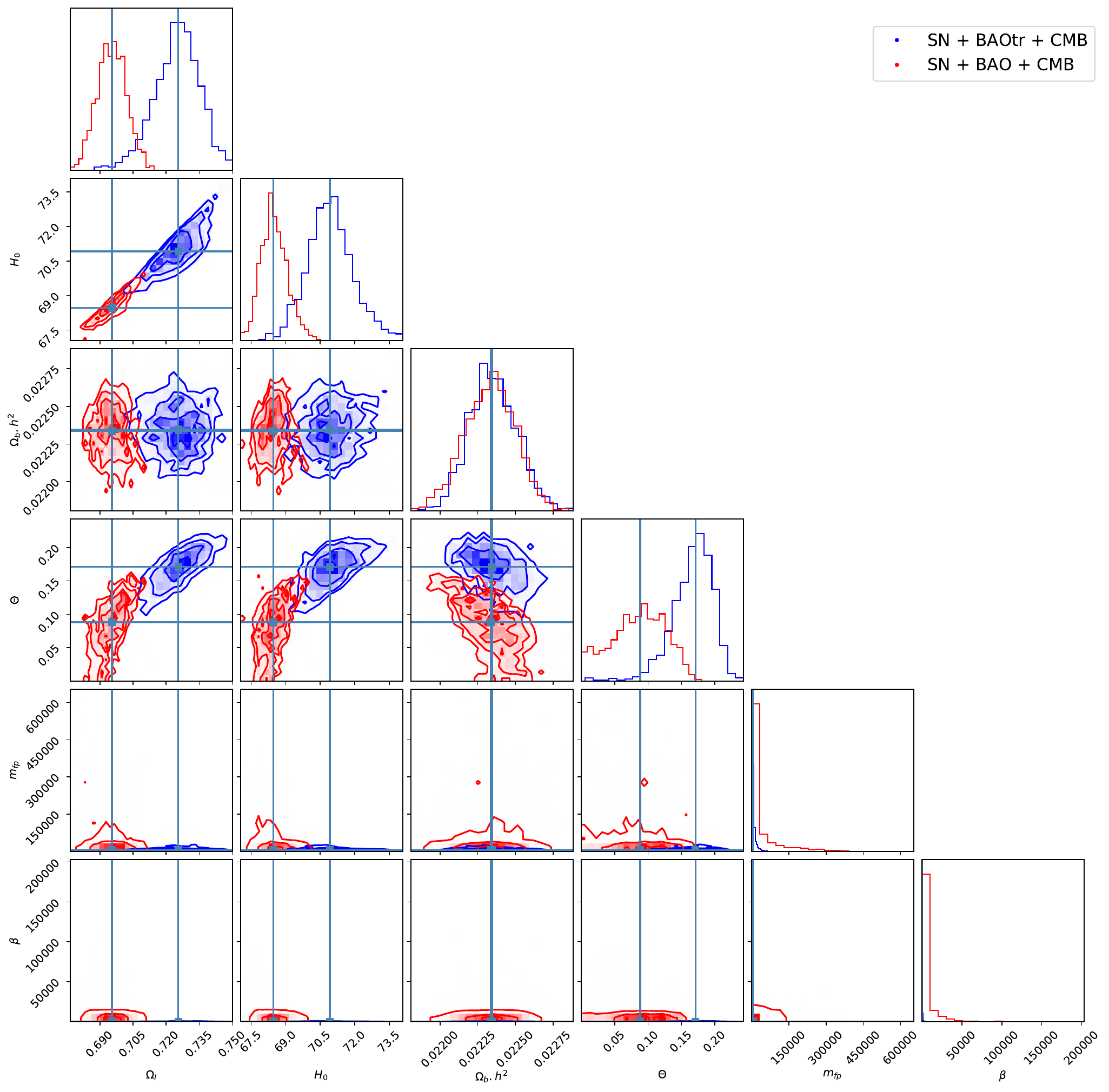}
    \caption{The $B_0B_1B_2B_3$-model. It is apparent that the inclusion of BAOtr instead of ordinary 3D BAO leads to a discernible increase in the estimated value of the Hubble constant $H_0$. In this four-parameter submodel, $\alpha$ is not a free parameter but becomes dependent on the remaining physical parameters, which is why it is not shown in the plot.}
    \label{fig:B0123}
\end{figure}

\begin{figure}[H]
    \centering
    \includegraphics[width=1\textwidth]{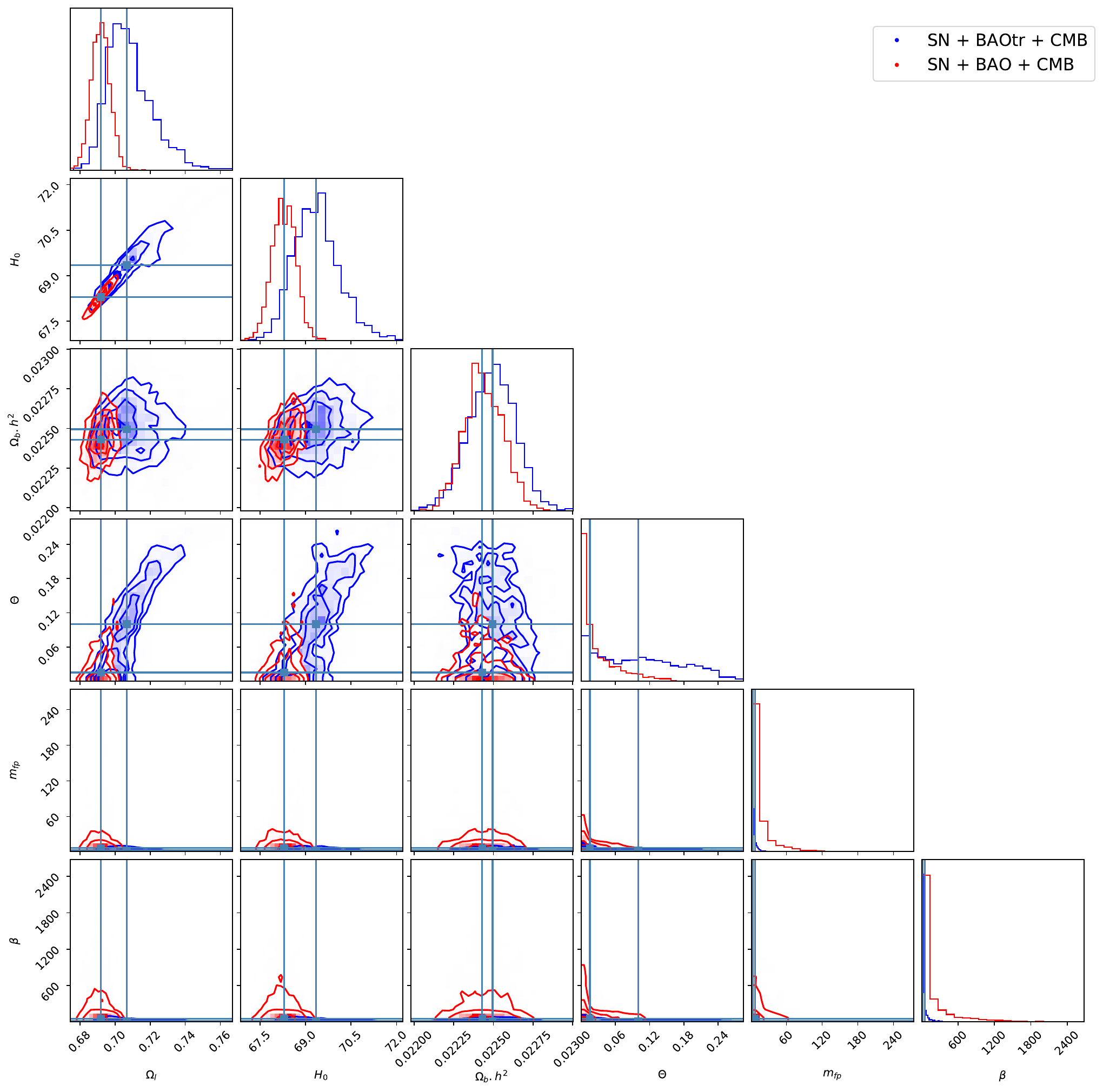}
    \caption{The $B_1B_2B_3B_4$-model. Incorporating BAOtr instead of ordinary 3D BAO results in a slight augmentation of the estimated value of the Hubble constant $H_0$. In this four-parameter submodel, $\alpha$ is not a free parameter but becomes dependent on the remaining physical parameters, which is why it is not shown in the plot.}
    \label{fig:B1234}
\end{figure}

\begin{figure}[H]
    \centering
    \includegraphics[width=1\textwidth]{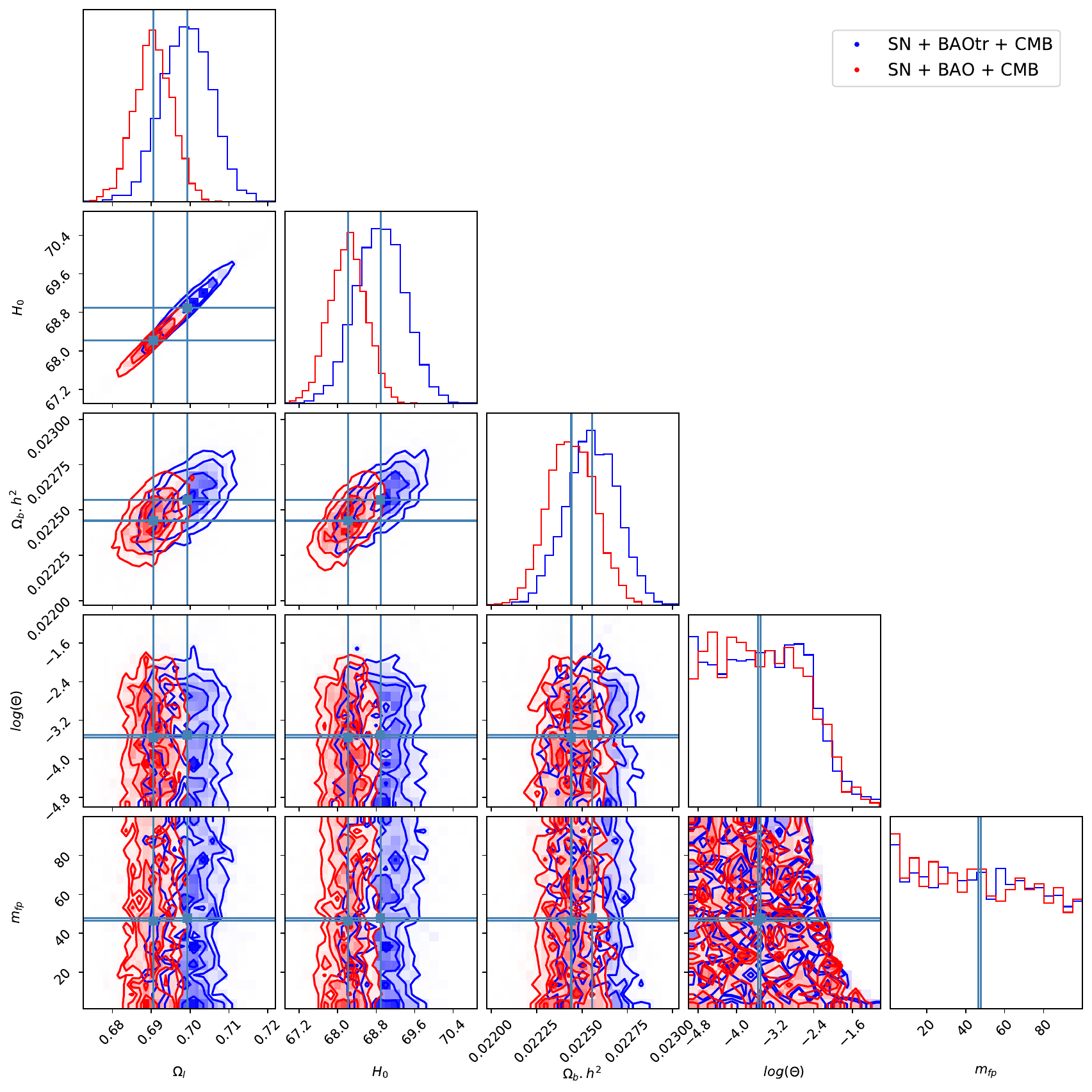}
    \caption{The $B_1B_2B_3$-model. The inclusion of BAOtr instead of ordinary 3D BAO leads to a slight increase in the estimated value of the Hubble constant $H_0$.In this three-parameter submodel, $\alpha$ and $\beta$ are not free parameters but become dependent on the remaining physical parameters, which is why they are not shown in the plot.}
    \label{fig:B123}
    
\end{figure}

\begin{figure}[H]
    \centering
    \includegraphics[width=0.6\textwidth]{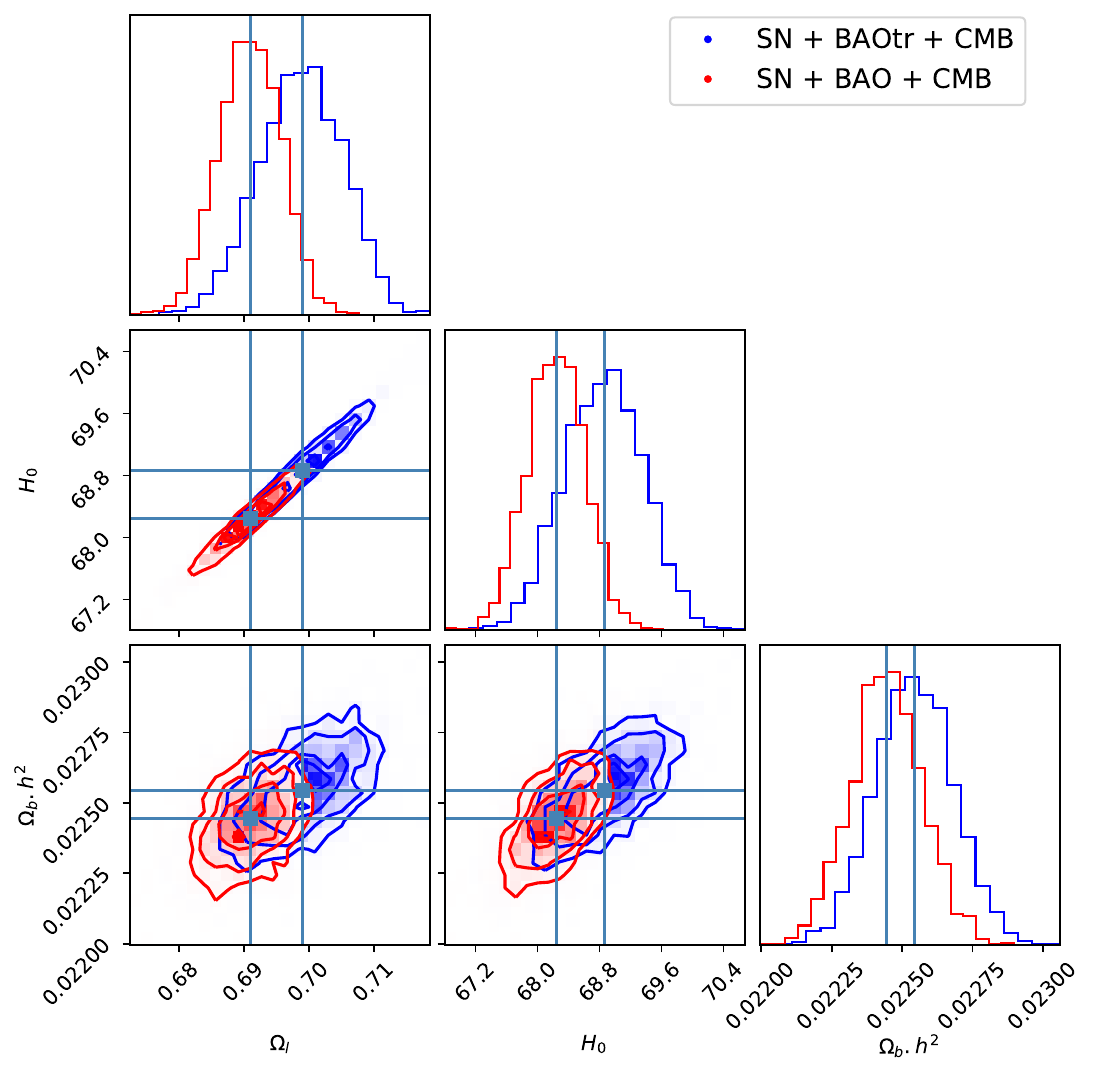}
    \caption{The corner plot shows the distribution of parameter values for a \lcdm model. It is apparent that including BAOtr instead of ordinary 3D BAO leads to a slight increase in $H_0$. However, the increment does not suffice to solve the Hubble tension and, moreover, for \lcdm, the usage of ordinary 3D BAO is warranted since the assumptions of the 3D BAO data reduction is compatible with such a cosmology.}
    \label{fig:LCDM}
\end{figure}

In Figs.~\ref{fig:B01234}--\ref{fig:B123} we show the full 2D marginalized CL contours for all bimetric submodels, as well as the \lcdm model model in Fig.~\ref{fig:LCDM}. As apparent in Fig.~\ref{fig:B01234}, the parameters $\alpha$ and $\beta$ are unconstrained when combining SN+BAO(tr)+CMB.
Concerning the $B_1 B_2 B_3$-model, displayed in Fig.~\ref{fig:B123}, we notice that the best-fit reduces to a \lcdm model, that is, $\theta \simeq 0$. This explains the peak in the likelihood for $\mfp$ towards higher values of $\mfp$. The reason is that for a $B_1 B_2 B_3$-model, the consistency constraints, discussed in Section~\ref{sec:ConsConstr}, enforce a small value for $\theta$ for large values of $\mfp$. This is discussed in detail in Section~3.2 in Ref.~\cite{Hogas:2021lns}. Moreover, in the $\theta \to 0$ limit, \lcdm is retained. So, a peak at large $\mfp$ is consistent with $\theta \simeq 0$ being the best-fit model. Already at $\mfp \sim 100$, it is required that $\theta \lesssim 0.1^\circ$ for a $B_1 B_2 B_3$-model which, for all practical purposes, represents a \lcdm model. Thus, it is not necessary to increase the parameter range in $\mfp$. 

\acknowledgments
Thanks to Armando Bernui and Rafael Nunes for valuable discussions on 2D BAO data and the $\Lambda_s$CDM model. Special thanks to Edvard Mörtsell for valuable comments on the manuscript and fruitful discussions which led to significant improvements of the analysis. Thanks to two anonymous reviewers for useful comments, improving the clarity of the manu\-script.

\bibliographystyle{JHEP}
\bibliography{References}

\end{document}